\documentclass[]{aastex631}

\usepackage{multirow}
\usepackage{amsmath}

\shorttitle{Shallow Weather, Deep Plumes and Vortices}
\graphicspath{{./}{Figures/}}

\begin{document} 

\title{Jupiter's Tropical Atmosphere: Shallow Weather, Deep Plumes and Vortices}

\correspondingauthor{Chris Moeckel}
\email{chris.moeckel@berkeley.edu}

\author[0000-0002-6293-1797]{Chris Moeckel}
\affiliation{University of California, Berkeley\\
Department of Earth and Planetary Science \\ 
307 McCone Hall, \\
Berkeley, CA 94720, USA}

\author[0000-0002-4278-3168]{Imke de Pater}
\affiliation{University of California, Berkeley \\ 
Department of Astronomy}

\author[0000-0001-9209-7716]{Bob Sault}
\affiliation{Independent}

\author[0000-0002-5344-820X]{Bryan Butler}
\affiliation{National Radio Astronomy Observatory}





\begin{abstract}
Towering storms, swirling clouds, and vortices are the cloud tops manifestation of complex weather systems shaping the atmosphere of Jupiter. We use observations from Juno's MicroWave Radiometer (MWR), the Very Large Array (VLA) and the Hubble Space Telescope (HST) to probe  for the first time the depth and impact of weather on Jupiter. 
We use ammonia, the main source of opacity at radio wavelengths on Jupiter, as the tracer for the weather by fitting ammonia anomalies to the MWR brightness temperature variations. 
We show that the majority of the weather on Jupiter is confined to regions where the clouds are forming. Both the South Equatorial Belt and the Equatorial Zone have surprisingly shallow weather systems (P $<$ 2 bar), and even in the North Equatorial Belt most of the ammonia variations is above the water condensation level (P $\sim$6 bar). 
This confirms that the water condensation layer plays a crucial role in controlling the dynamics and the weather on Jupiter.  However, the shallow nature of the weather cannot explain the deep-seated depletion down to 30 bars that the Juno mission has revealed. We do find three features, however, that extend below the water condensation layer: a vortex in the northern hemisphere reaching down to 30 bars, an ammonia plume down to 20-30 bars, and the signature of precipitation down to 20 bars. 

This work confirms that an interplay of large-scale processes (vortices, plumes) and small-scale processes (storms) are responsible for shaping the atmospheric makeup of Jupiter. 

\end{abstract}



\section{Introduction} \label{sec:intro}
The Juno mission has challenged our understanding of the troposphere of Jupiter, by uncovering a distribution of material within the atmosphere that is inconsistent with our understanding of the dynamics. Before the Juno mission, the main driver of atmospheric distribution was assumed to be the global circulation, with Hadley and Ferrel cells distributing energy and material in the atmosphere \citep[see, e.g., the review by][]{depater2023}. More recently, the circulation has been interpreted by a 2-layer shallow circulation \citep{Ingersoll2000,Showman2005,Fletcher2009,Fletcher2020}, where the upper circulation cell extends from the tropopause to the layers at which the water condenses (P $\sim$6 bar). This circulation was challenged by the realization that ammonia \citep{Li2017,Moeckel2023} and temperature effects \citep{Li2024,Moeckel2024} can be traced to much higher pressures (P $\geq$20 bar). Instead, recent theoretical work \citep{Guillot2020} and observations \citep{Moeckel2024} support the idea that processes on the smallest scales could potentially explain the distribution by depleting the upper troposphere and enriching the lower troposphere through storms. 

For this, processes at various scales have to interact to create these conditions. It requires a large-scale suppression of convection in order to build up convective available potential energy (CAPE) through mass loading in the atmosphere, which is then violently released on intermittent time scales to form large-scale storms and localized thunderstorms \citep[e.g.,][]{Showman2005}. Inside these storms, the condensation and precipitation of water can be delayed, supercooling the water to a point where ammonia can much more effectively dissolve into water \citep{Guillot2020}. Due to the supercooled nature of the water, the precipitates that form in this process can rapidly grow and obtain significant falling speeds, to reach pressures and temperatures much beyond their melting and evaporation points. This impact of storms on the middle ($\sim$0.7--6 bar) and lower ($P > 6$ bar) troposphere has been observed on both Jupiter \citep{Moeckel2024} and Saturn \citep{Li2023}, and could also play an important role for the ice giants \citep{Guillot2021}. 

In this paper we focus on intermediate scales - the weather patterns in the tropics of Jupiter, where we define ``tropics'' as the region between latitudes $\sim$20$^\circ$S and $\sim$20$^\circ$N. The turbulent cloud tops that dominate the tropical region of Jupiter offer us a glimpse of the dynamics that constantly shape the atmosphere.  While we have a good understanding of the depths to which large-scale outbreaks \citep{dePater2019,Moeckel2024} and vortices \citep{Bolton2021} can affect the atmosphere, we lack an understanding of the role of these weather systems in affecting the global atmosphere. In this work, we explore the link between the dynamics that play out at the cloud tops and its effect on the atmosphere below the clouds.
 
We combine HST observations with concurrent radio observations obtained with the MicroWave Radiometer (MWR) on board of the Juno spacecraft to map the three-dimensional structure of Jupiter's ammonia distribution, the main source of opacity at radio wavelengths. We use simultaneously obtained observations with the Very Large Array (VLA) to provide context and generalize these findings. We focus our analysis on the tropical region of Jupiter, where visible and radio data show the majority of atmospheric activity. We first describe these three data sets (Section \ref{sec:data}). In Section \ref{sec:method} we explain our methodology to retrieve the three-dimensional weather from the Juno observations. Section \ref{sec:results} shows our results, followed by a comparison of the Juno and VLA observations and generalizing these findings to the global atmosphere (Section \ref{sec:discussion}).  Section \ref{sec:conclusions} concludes the paper.

\section{Data} \label{sec:data}
Perijove 19 (PJ19) was dedicated to study the longitudinal structure of Jupiter, by aligning the spacecraft's spin axis with the velocity vector, allowing the atmospheric instruments to scan the planet in an east-west manner. This specific focus on the longitudinal structure of the spacecraft motivated many scientists to support the campaign with (near)-simultaneous earth-based observations across a wide range of frequencies. 

\subsection{VLA data}
We observed Jupiter with the VLA across 3 frequency bands in the B-configuration during Juno's closest approach to provide a global snapshot of the Jovian sub-cloud structure down to pressures of $\sim$ 4 bar (see Figure \ref{fig:PJ19WF}). To capture a global map for each frequency, we observed Jupiter over 4 nights, each night covering a Jovian hemisphere. The first two nights were used to observe Jupiter in X-band (8-12 GHz), interleaved with Ku-band (12-18 GHz); the two remaining nights were used for K-band (18-26 GHz). We detail the relevant details for each observation in Table \ref{tab:obsdata}. Each night was processed using the standard VLA CASA pipeline (Common Astronomy Software Applications) and then further processed in CASA (version 5.4.1-32).  All four nights used the same amplitude and phase calibrator (3C286 and J1726-2258).

We first averaged the observations in time and frequency (10s, 16MHz) to increase the signal-to-noise ratio and reduce the size of the dataset for processing. We analyzed the difference between the expected and observed amplitudes as a proxy for the noise and interference in the system for each antenna pair. We flagged observations that showed spurious signals that were not picked up by the automatic radio frequency interference task in the pipeline.  Once the data were flagged, we combined the two nights into a single combined dataset for the iterative inversion process. We split each night into two frequency bands for the self-calibration: The first two nights were split into X-band (8-12 GHz) and Ku-band (12-18 GHz) along their designated frequency band allocations, while the third and fourth nights were split in the middle (18-22 GHz, 22-26 GHz). \\ 
Variations in the atmospheric conditions above the antennas affect the phase of the incoming radio waves and increase the noise of the observations. The phase of the incoming radio waves is used in the Fourier transform to determine the location of the source on the sky. Therefore, phase delays blur the brightness distribution on the sky. Since we have a good understanding of the large-scale structure on the planet, we can correct for the phase-delays by correlating the signal between the various antenna pairs in a process called self-calibration. For the self-calibration procedure we build a model atmosphere that mimics the position and shape of Jupiter based on the JPL horizons ephemeris\footnote{\url{https://ssd.jpl.nasa.gov/horizons/}} and observed flux densities from previous observations \citep{dePater2019b}. Due to the high resolution of the observations and the wide spectral coverage, we included the zones and belts and their dependencies on the frequency in our model atmosphere. Within CASA we Fourier transform the model into the {uv}-plane, and performed two iterations of self-calibration. With increasing iterations we decrease the duration over which we obtain the phase delays starting with a solution interval of 5 min, and decreasing it to 2 min in the second iteration. These intervals represent the duration of the atmospheric fluctuations that we average over. We found that at shorter time intervals the signal was not strong enough to further solve for phase-delays. This concluded all the calibration efforts in CASA and we exported the calibrated dataset for the deprojection step in MIRIAD (Multichannel Image Reconstruction, Image Analysis and Display) \citep{Sault1995}. \\ 
We noticed that each observation of Jupiter appeared to be shifted from its ephemeris. The presence of these shifts probably points to unresolved phase-delays affecting the observations, perhaps introduced by the self-calibration process. We, therefore, move the center of the data until they correspond well with the expected location of Jupiter for the given epoch. For the deprojection we remove the bulk of the emission by subtracting a limb-darkened disk and map the residuals using the method developed by \citet{sault2004} to obtain global radio-residual maps. We produce maps that make use of the full frequency coverage for each band, where the increased frequency coverage fills in more of the \textit{uv}-space increasing the fidelity of the Fourier transform. 
The radio residuals highlight the atmospheric structure as seen from Earth for a rotating planet. 


The emission angle $\theta_e$, the angle between the local zenith and the radio observatory, varies across latitude due to the shape of the planet (cos$\theta$) and varies with longitude due to the observation geometry during the observations, \textbf{ albeit the longitudinal variation is relatively small \citep{sault2004}}. 


Without prior knowledge about the atmosphere, we cannot correct for this effect, since the limb-darkening is dependent on the distribution of radio-absorbing gases in the atmosphere. \textbf{ Therefore, the VLA maps should be seen as lower limits to variability \citep{Hardesty2025}. }  

\subsection{HST data}
Hubble Space Telescope observations (GO-\#15665, PI: de Pater) were taken as part of the VLA observations campaign in support of PJ19. In total, they cover 4 rotations of Jupiter. The data are publicly available in the Mikulski Archive for Space Telescopes\footnote{\url{https://archive.stsci.edu/hlsp/wfcj}} and were published by \cite{Wong2020}; we refer the reader to this paper for the description of the data reduction. During the campaign HST took images at 225 nm, 275 nm, 395 nm, 631 nm, 727 nm, and 889 nm, however we only use the blue 395 nm and red 631 nm to provide context for the radio observations. 

\subsection{Juno/MWR}
Juno hosts a set of six radiometers operating between 0.5 and 22 GHz, referred to as Channel 1 through 6 (C1-C6), to study the troposphere of Jupiter \citep{Janssen2017}. The six radiometers record the beam-convolved brightness temperature, that is the target brightness temperature distribution convolved with the instrument's sensitivity pattern (the beam). The MWR beams sweep across the atmosphere as the satellite spins with 2 rpm, obtaining multiple measurements for a given location on the planet at different emission angles. The rotation vector of the spacecraft determines the number of overlapping observations. As mentioned, PJ19 was optimized for longitudinal coverage.

For our data reduction we obtain the antenna temperatures from the Planetary Data System \footnote{\url{https://pds.nasa.gov/}} and first fit the nadir brightness temperature and limb-darkening coefficient to all the observations based on a parametric relationship between the brightness temperature and the emission angle \citep{Moeckel2023}. We choose a 2$^\circ$ bin size for this dataset, that is we consider all observations for which the boresight fell within the bin size to fit the parameters. Based on this we obtain zonal brightness temperature and limb-darkening coefficient along with their uncertainties. 
We use this average as the starting point for the deconvolution that maps sub-beam size structure across the planet \citep{Moeckel2024}. In an iterative approach, we convolve our best understanding of the atmospheric distribution with the beam pattern for each observation and update the maps based on the mismatch between observations and the best fit. Since we are convolving the brightness temperature structure with viewing geometry, we lose information about the limb-darkening coefficients, reducing our observables from 12 independent measurements (six frequencies sampling the nadir brightness temperature and limb-darkening coefficient) to just the six nadir brightness temperatures. We remove measurements where the beam is only partially on the planet to decrease the synchrotron contamination \citep{Moeckel2023}. The larger beamsize of C1 and C2 causes us to throw out more measurements resulting in a much narrower domain, compared to the upper channels and limits longitude/latitude extent over which we have full frequency coverage.

We used all available measurements with their varying viewing geometries to construct maps. Closest to the center of the map we have the best resolution, steadily decreasing towards the edge of the domain due to the large footprint. The measurements at the lower frequencies closest to the edge of the domain appear to be affected by synchrotron radiation, despite limiting the beams we are using for the deconvolution. Instead of restricting the emission angle coverage we decided to remove the synchrotron radiation by running a moving median filter on the brightness temperature residuals for C1 -  C3 (see Appendix A).

\begin{figure}
\centering
\includegraphics[width=0.66\textwidth]{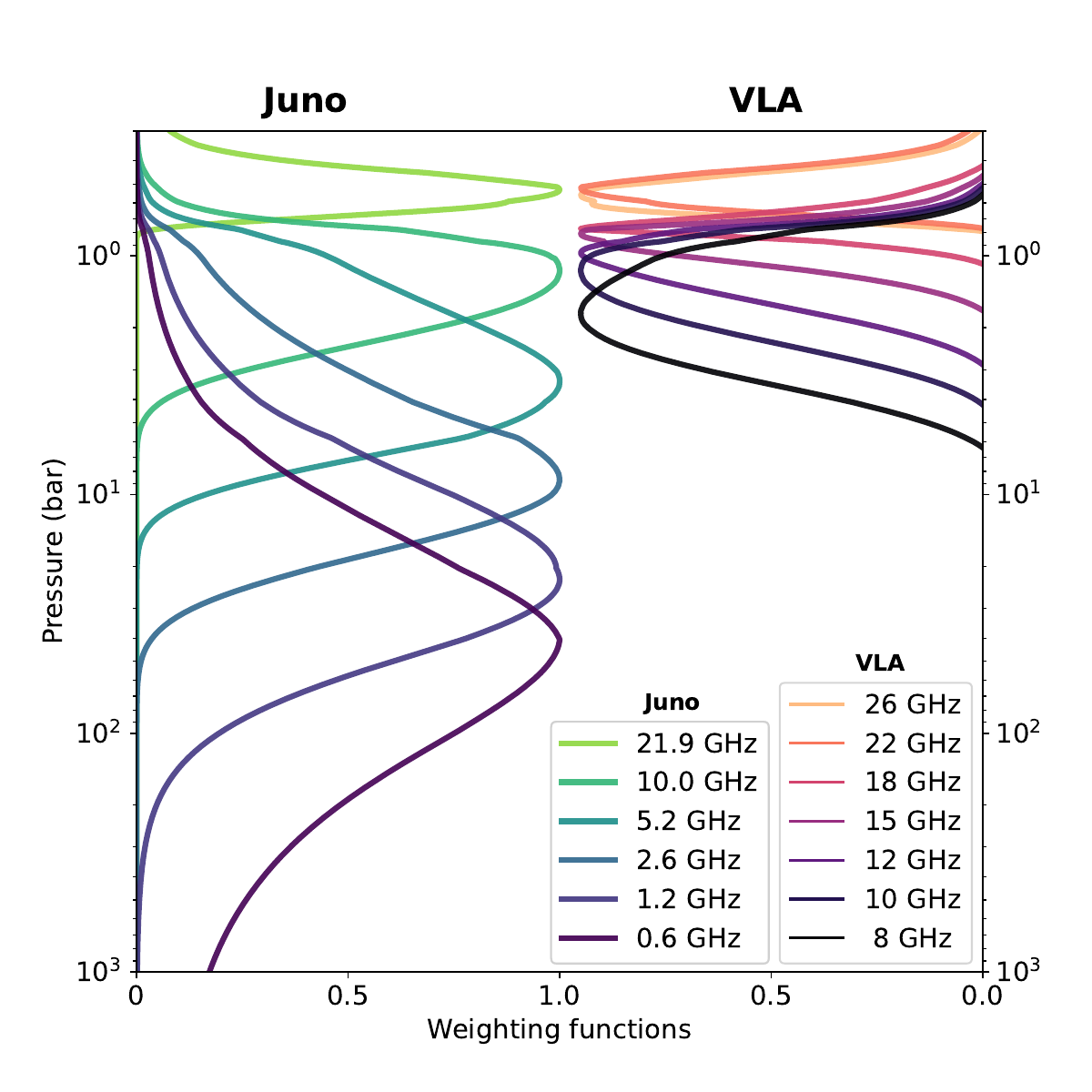}
\caption{Nadir probing weighting functions that indicate\textbf{s the pressure range that the thermal emission is integrated over} for a given frequencies. The Juno frequencies are presented on the left hand side, while the VLA frequencies are shown on the right hand side with lower frequencies probing deeper into the planet. The weighting functions are based on the mean tropical atmosphere of Jupiter that we retrieved for PJ19 (see Figure \ref{fig:PJ19-NH3-ZonalMap}) coupled to a dry adiabat \citep{Moeckel2023}.}
\label{fig:PJ19WF}
\end{figure}

\begin{table}[]
\caption{Overview of the observational coverage that is presented in this work with their corresponding program codes, their frequency coverage, the timing of the observations and their approximate resolution on the planet. We combine HST observations (top third) with VLA observations (middle third) and Juno MWR observations (lower third). }
\label{tab:obs}
\begin{tabular}{|c|c|c|c|c|c|}
\hline
                                                                                                      & \textbf{Label} & \textbf{$\lambda$ (nm)}        & \textbf{Date}   & \textbf{Time (UTC)}                  & \textbf{Resolution (km)}      \\ \hline
\multirow{8}{*}{\rotatebox[origin=c]{90}{\begin{tabular}[c]{@{}c@{}}Hubble Space Telescope \\ GO-\#15665 \end{tabular}}}  & HST - blue     & 395                             & \multirow{2}{*}{2019-04-06} & 10:36:19                             & \multirow{8}{*}{$\sim$180} \\ \cline{2-3} \cline{5-5}
                                                                                                      & HST - red      & 631                             &                             & 10:39:57                             &                               \\ \cline{2-5}
                                                                                                      & HST - blue     & 395                             & \multirow{2}{*}{2019-04-07} & 10:26:18                             &                               \\ \cline{2-3} \cline{5-5}
                                                                                                      & HST - red      & 631                             &                             & 10:29:56                             &                               \\ \cline{2-5}
                                                                                                      & HST - blue     & 395                             & \multirow{2}{*}{2019-04-08} & 10:16:15                             &                               \\ \cline{2-3} \cline{5-5}
                                                                                                      & HST - red      & 631                             &                             & 10:19:53                             &                               \\ \cline{2-5}
                                                                                                      & HST - blue     & 395                             & \multirow{2}{*}{2019-04-09} & 10:01:56                             &                               \\ \cline{2-3} \cline{5-5}
                                                                                                      & HST - red      & 631                             &                             & 09:58:31                             &                               \\ \hline
                                                                                                      & \textbf{Label} & \textbf{$\nu$ (GHz) } & \textbf{Observation date}   & \textbf{Time (UTC)}                  & \textbf{Resolution (km)}      \\ \hline
\multirow{6}{*}{\rotatebox[origin=c]{90}{\begin{tabular}[c]{@{}c@{}}Very Large Array \\ VLA ID: 19A-001\end{tabular}}}    & X-band   & 8 - 12     & \multirow{2}{*}{2019-04-06} & 08:58:01 - 14:27:06                  & $\sim$4000                    \\ \cline{2-3} \cline{5-6} 
                                                                                                      & Ku-band  & 12 - 18     &                             & 08:58:01 - 14:27:06                  & $\sim$2500                    \\ \cline{2-6} 
                                                                                                      & X-band   & 8 - 12      & \multirow{2}{*}{2019-04-07} & 08:54:05 - 14:23:09                  & $\sim$4000                    \\ \cline{2-3} \cline{5-6} 
                                                                                                      & Ku-band  & 12 - 18       &                             & 08:54:05 - 14:23:09                  & $\sim$2500                    \\ \cline{2-6} 
                                                                                                      &  K-band   & 18 - 26     & 2019-04-08                  & 08:50:09 - 14:19:14.0                & $\sim$1500                    \\ \cline{2-6} 
                                                                                                      &  K-band   & 18 - 26      & 2019-04-09                  & 08:46:13 - 14:15:18                  & $\sim$1500                    \\ \hline
                                                                                                      & \textbf{Label} & \textbf{$\nu$ (GHz)} & \textbf{Observation date}   & \textbf{Time (UTC)}                  & \textbf{Resolution (km)}      \\ \hline
\multirow{6}{*}{\rotatebox[origin=c]{90}{\begin{tabular}[c]{@{}c@{}}Juno MWR \\ PJ19 - Crosstrack\end{tabular}}}                                                  & C1             & 0.6                 & \multirow{6}{*}{2019-04-06} & \multirow{6}{*}{12:05:18 - 12:31:18} & $\sim$2000                    \\ \cline{2-3} \cline{6-6} 
                                                                                                      & C2             & 1.248                 &                             &                                      & $\sim$2000                    \\ \cline{2-3} \cline{6-6} 
                                                                                                      & C3             & 2.597           &                             &                                      & $\sim$1000                    \\ \cline{2-3} \cline{6-6} 
                                                                                                      & C4             & 5.215             &                             &                                      & $\sim$1000                    \\ \cline{2-3} \cline{6-6} 
                                                                                                      & C5             & 10.0                &                             &                                      & $\sim$1000                    \\ \cline{2-3} \cline{6-6} 
                                                                                                      & C6             & 21.9                &                             &                                      & $\sim$1000                    \\ \hline
\end{tabular}
\label{tab:obsdata}
\end{table}

\subsection{Radio Maps}

\begin{figure}
\centering
\includegraphics[width=0.85\textwidth]{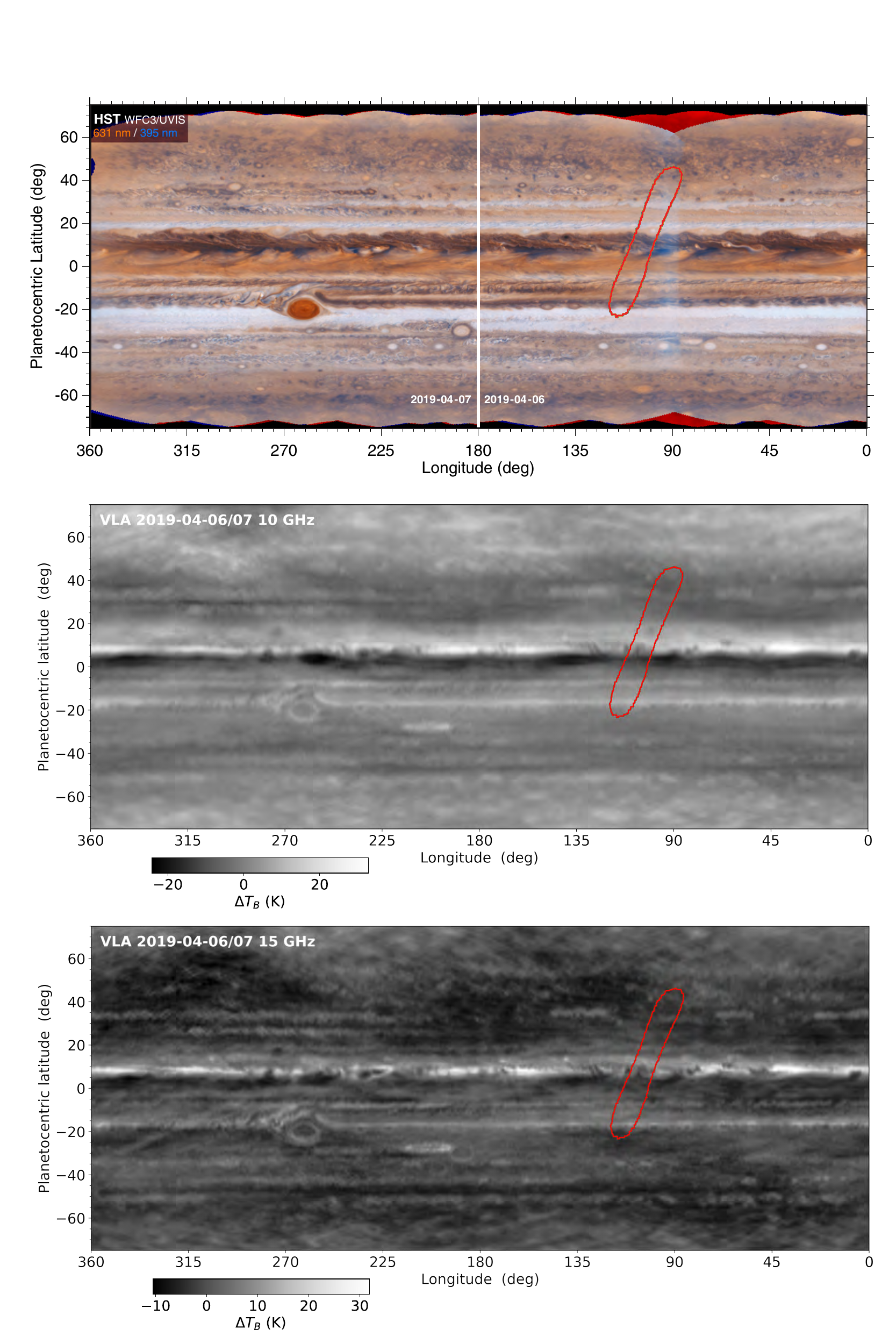}
\caption{HST and VLA X- and Ku-band observations on April 6-7. The observing geometry from Earth only allowed hemispherical coverage of the planet, so that the two hemisphere are separated by 24 hours, or 2.5 Jovian rotation. The top row shows the red and blue filter composite of the HST observations, the middle row shows the VLA X-band observations centered around 10 GHz with a 4 GHz bandwidth, and the lower row shows the VLA Ku-band observations centered around 15 GHz with a 6 GHz bandwidth. A Gaussian filter was used for the Ku band data to remove any non-zonal large scale structure (see Appendix \ref{sec:appB}). The red outline indicates where we have simultaneous coverage across all 6 Juno frequencies. }
\label{fig:xku-maps}
\end{figure}

In Figure \ref{fig:xku-maps} we show the observations from April 6 and 7 that coincided with Juno's PJ19. The three panels show the HST observations, the VLA X-band observations, and the VLA Ku-band observations, from top to bottom. The outline indicates the fraction of the planet where we have full frequency coverage of all 6 channels with MWR. However, Juno's lower frequency measurements are able to provide information to a much deeper part of the Jovian atmosphere (see Figure \ref{fig:PJ19WF}). The VLA X-band maps, probing deepest from all the VLA maps, show that Juno passed just west of an equatorial ammonia plume (90$^\circ$W, 6$^\circ$N, all latitudes are planetocentric in this paper, see also Figure \ref{fig:PJ19-TropicalZoomMap} for a zoomed in version), as seen by extremely low brightness temperatures compared to their surroundings. Furthermore, the VLA maps show that the variability within the PJ19 region is typical for that part of the atmosphere. In the southern hemisphere, the Juno observations span two very narrow bright bands in the X-band map, whereas in the northern hemisphere the vortex structure at 17$^\circ$N appears as a cold brightness temperature anomaly. 
Probing slightly higher in the atmosphere, the Ku-band maps show much more variability in the atmosphere than the X-band maps. The ammonia plume anomaly just east of the flyby is less pronounced, but the slightly higher resolution of the Ku-band observations shows that Juno covered an interesting region with significant brightness temperature variations at the EZ-NEB interface. 

\begin{figure}
\centering
\includegraphics[width=0.85\textwidth]{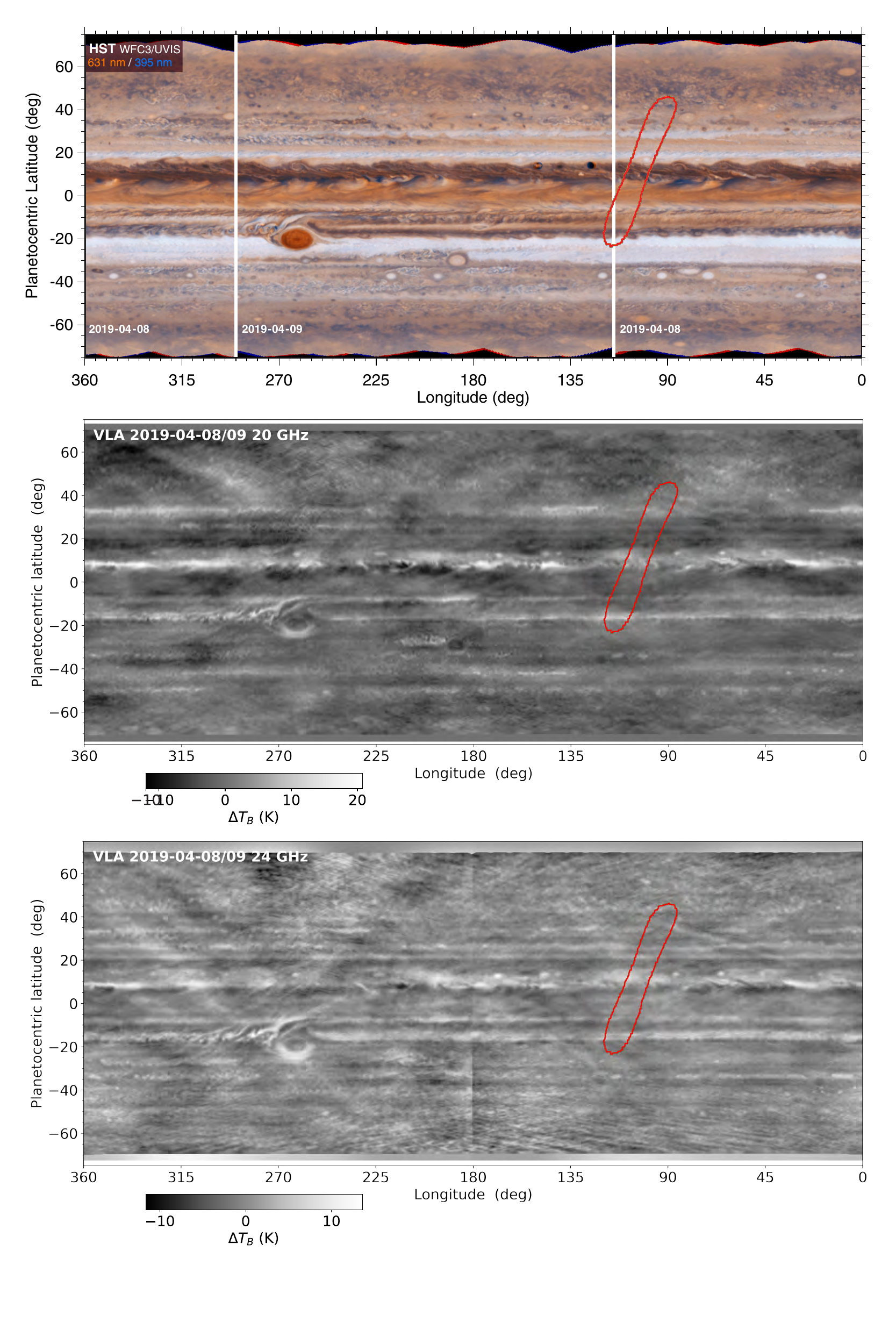}
\caption{Same as Figure \ref{fig:xku-maps} but for the observations on April 8-9. The middle row shows the VLA K-band observations around 20GHz with a 4 GHz wide bandwidth, and the lower row shows the K-band observations around 24 GHz with a 4 GHz bandwidth. This lower panel was heavily filtered, and should be considered qualitative. }
\label{fig:k-maps}
\end{figure}

In Figure \ref{fig:k-maps} we show the observations for the next two days. The K-band observations probe the pressures around the ammonia ice cloud, i.e., the interface between the middle and the upper troposphere. The middle panel (centered around 20 GHz) shows that Juno observed a pretty calm part of the atmosphere at these pressures. The lower panel in Figure \ref{fig:k-maps} is centered at 24 GHz, close to the center of the ammonia absorption band (23.5 GHz). At this frequency, the observations probe the highest altitudes in the atmosphere. These observations, being close to the water absorption band, were the most challenging to reduce, likely due to variability in the water content in Earth's atmosphere. There are significant non-zonal features that appear to be artifacts. Due to heavy filtering, we consider these maps only qualitative. The overall variability decreases in the maps as the observations probe altitudes at and above the NH$_3$ condensation level. The regions in the maps that Juno sampled appear mostly uniform at these altitudes.

\begin{figure}
\centering
\includegraphics[width=0.85\textwidth]{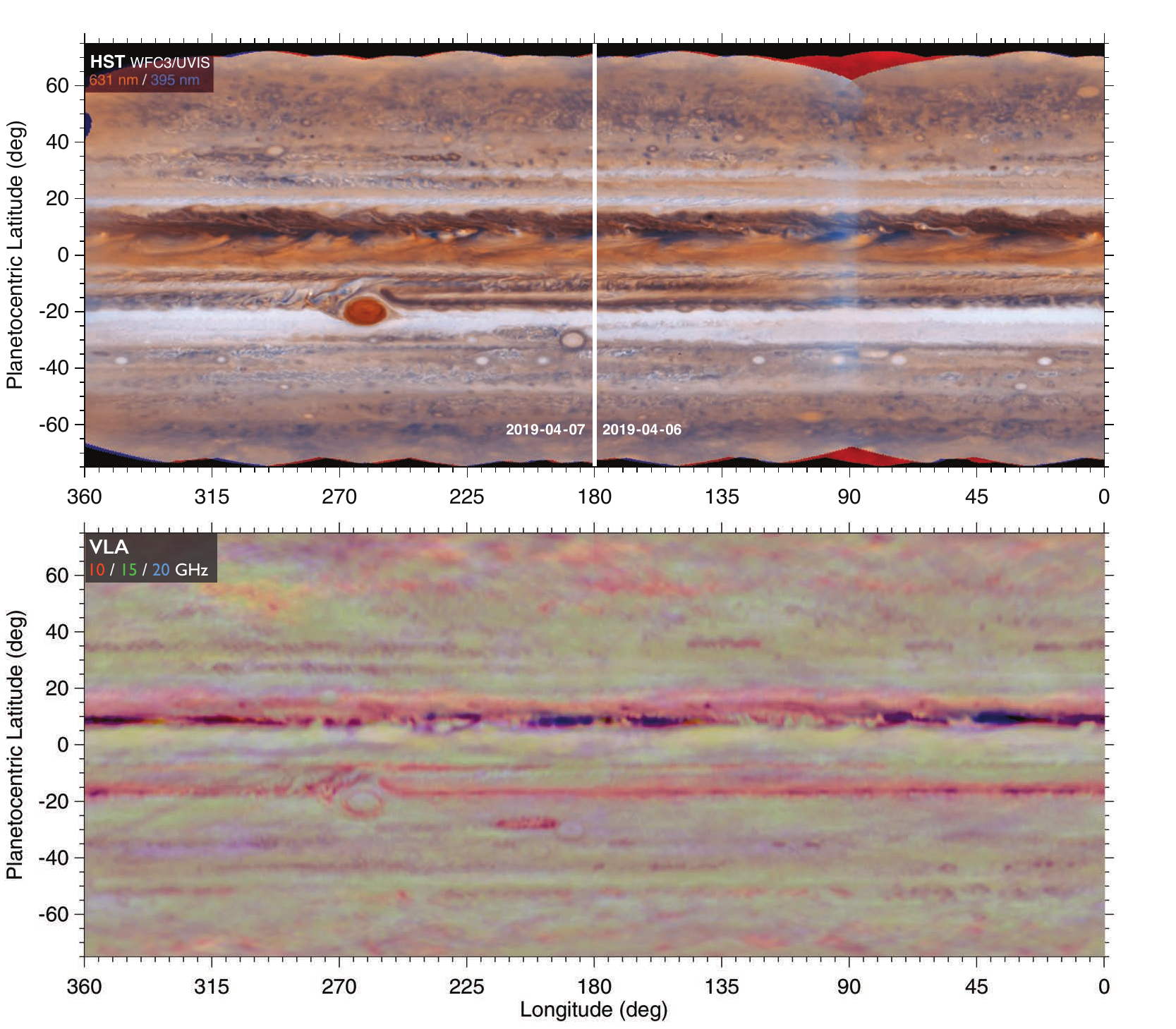}
\caption{HST and VLA observations taken during Juno's PJ19 encounter. The top panel shows the HST observations, while the lower contains a composite made of the three frequency bands corresponding to three color channels: red and yellow (X-band/10 GHz) show warm and cold brightness temperatures deepest in the planet observable with the VLA, green/white colors (Ku-band/15-GHz) show the structure under the ammonia ice clouds, and blue colors (K-band/20-GHz) probe the pressures around the ammonia ice clouds. }
\label{fig:radio-composite}
\end{figure}

For visualization purposes only, we combine three of the four maps into a color composite image where the different colors indicate the depths from which we receive the signal. The red colors in Figure \ref{fig:radio-composite} correspond to radio-warm structure that has a deep root. Yellow corresponds to structure that is deep and cold in the planet, like the vortices at the northern edge of the NEB or the plumes at the northern edge of the EZ. The whiter/greenish tones correspond to shallow structure extending just below the ammonia ice clouds, while blue/purple correspond to structure in and above the ammonia clouds.

\begin{figure}
\centering
\includegraphics[width=\textwidth]{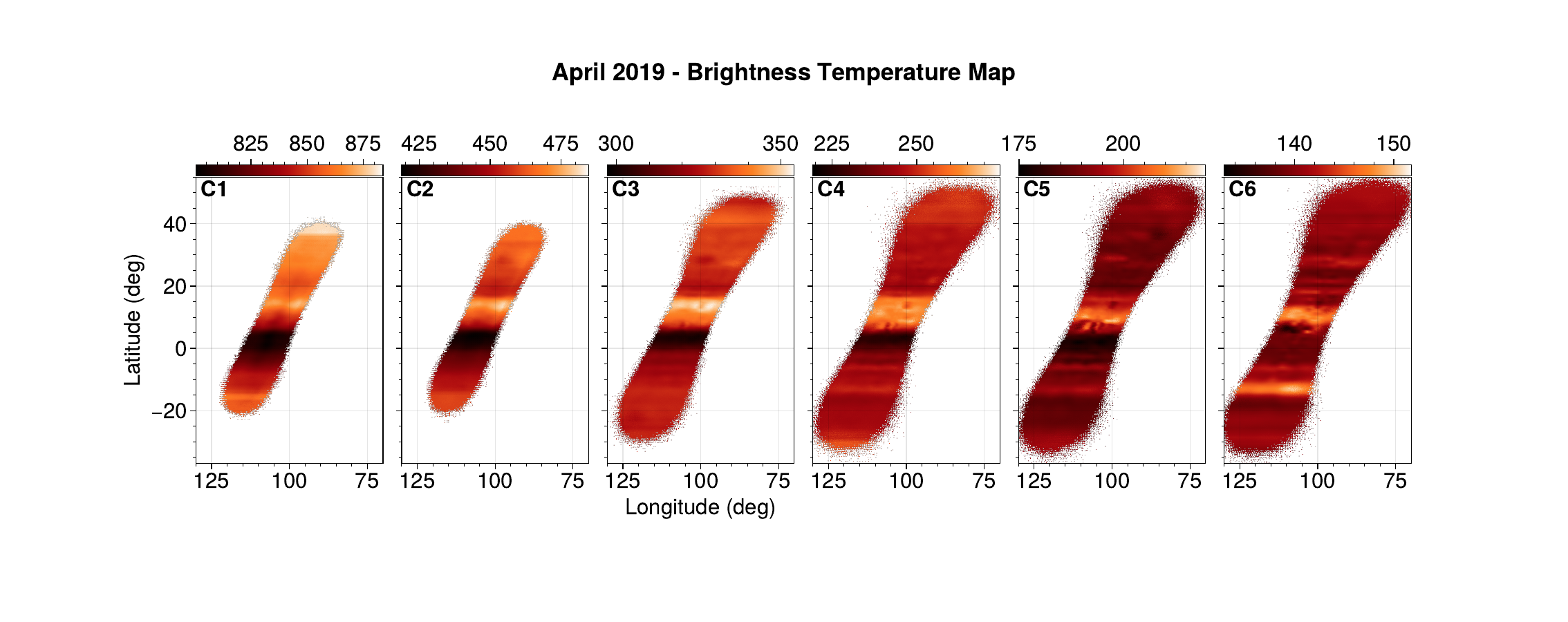}
\caption{Deconvolved PJ19 Juno observations that map out the nadir brightness temperature distribution for the 6 channels. The channel numbers are increasing left to right. The colorbar for each panel is appended to the top of the plot. The resolution is set by the size of the individual beams and the number of overlapping observations \citep{Moeckel2024}. The lower frequencies cover less area due to the larger beamsize and therefore dictate the region over which we have full frequency coverage. These maps were independently produced, analyzed, and published by \citep{Bolton2021}.}
\label{fig:juno-maps}
\end{figure}

The Juno deconvolved maps in Figure \ref{fig:juno-maps} show the nadir brightness temperature distribution for the 6 channels. Interpretation of the brightness temperatures alone is complicated by the fact that the weighting functions cover a wide range of pressures. Features visible across multiple frequency bands, however, are good indications of the vertical extent of these features. These maps reproduce the structure of the maps published before \citep{Bolton2021}. The most significant feature in these maps are the two large-scale features, an anticyclone and a barge in the northern hemisphere, that are visible across most of the frequencies.

\section{Methodology} \label{sec:method}
The zonally averaged Juno data (brightness temperature and limb-darkening) allows us to constrain Jupiter's tropical region to a much higher accuracy and greater depth than with the VLA observations, albeit with a narrow longitudinal focus dictated by the orbit. Combining the VLA and Juno data in our radiative transfer retrievals would require us to downconvolve the Juno observations to the VLA resolution. Instead, we use the Juno observations to build up a 3D map of the temperature and NH$_3$ distribution in the tropical region of the atmosphere, and compare the results directly to the VLA observations. 

\begin{figure}
\centering
\includegraphics[width=\textwidth]{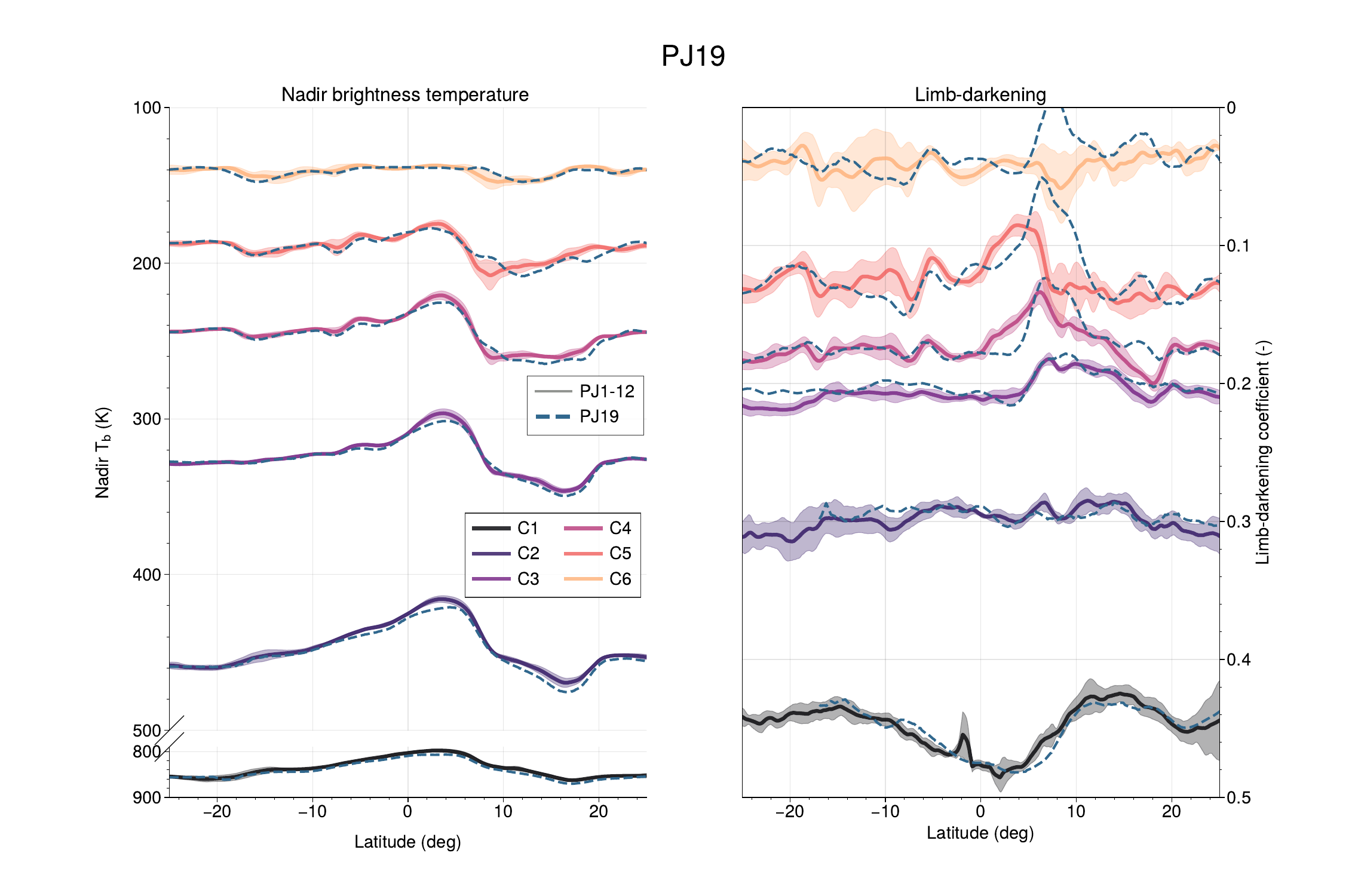}
\caption{PJ19 observations compared to the mean atmosphere composed of the average of PJ1--12 \citep{Moeckel2023}. The PJ19 observations are used to build up the reference atmosphere for our analysis. }
\label{fig:juno-observations}
\end{figure}

\begin{figure}
\centering
\includegraphics[width=\textwidth]{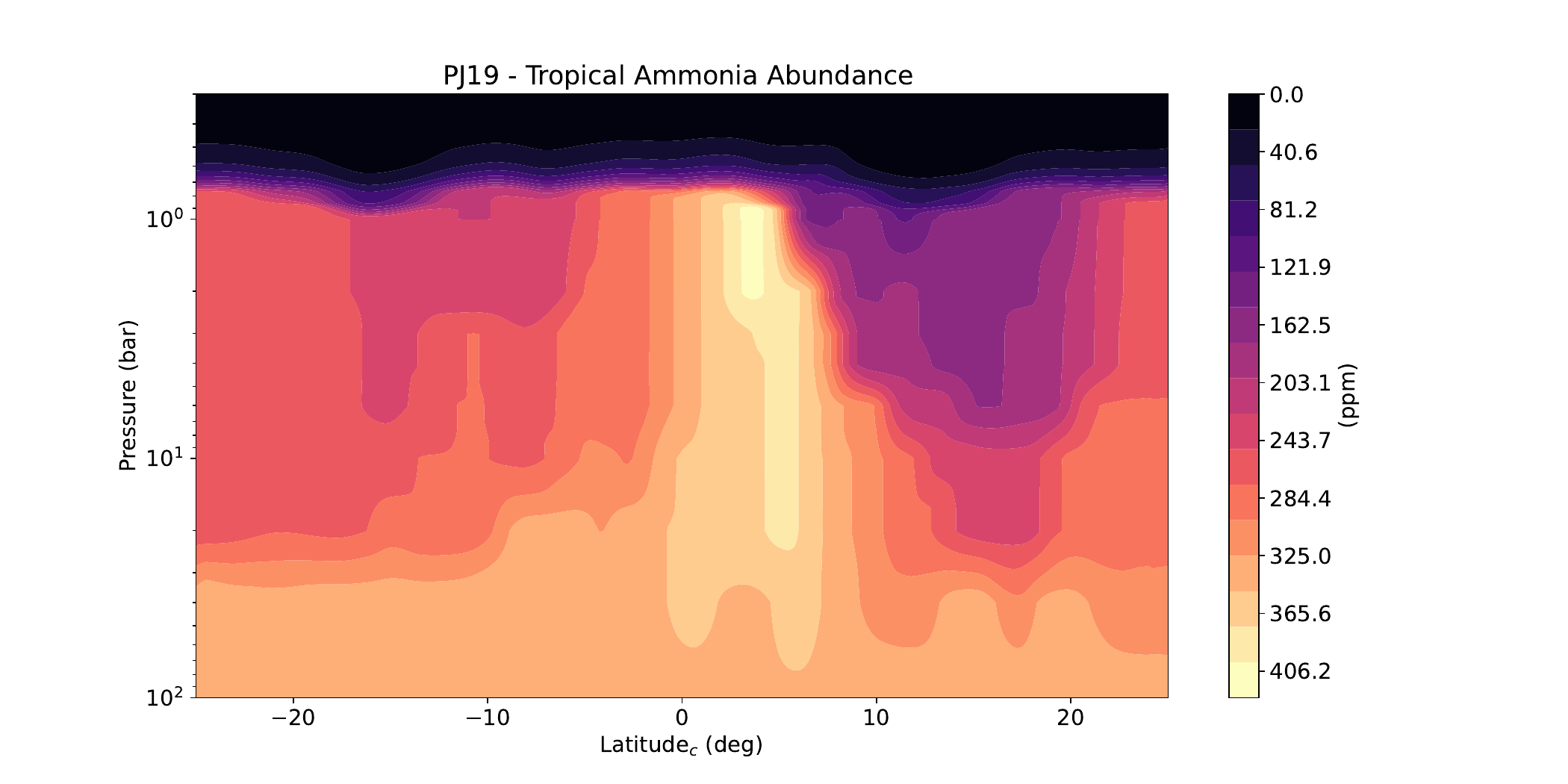}
\caption{Retrieved ammonia abundance based on all zonally averaged Juno observables for PJ19 (see Equation \ref{eq:C1}). We median filtered this map for visualization purposes only to remove small scale fluctuations. } 
\label{fig:PJ19-NH3-ZonalMap}
\end{figure}

Our goal is to map variations in the three-dimensional ammonia distribution. Since we are less interested in the absolute abundances,  we utilize a multi-step method to obtain the ammonia distribution in Jupiter's atmosphere. We first constrain the ammonia abundance by fitting a model atmosphere that matches the zonally averaged Juno PJ19 observations; based on both the nadir brightness temperature and the limb-darkening coefficient \citep{Moeckel2023}. Figure \ref{fig:juno-observations} shows the PJ19 observables compared to the mean of the first 12 orbits (from \citet{Moeckel2023}), including their standard deviations. The PJ19 signal in the southern hemisphere is within the nominal atmosphere, while the atmosphere in the northern hemisphere shows more variations. 
We use radiobear \footnote{\url{https://github.com/david-deboer/radiobear}} \citep{dePater2005,dePater2014,dePater2019b,Moeckel2023} to model the effective brightness temperature for a given ammonia and/or temperature distribution in the atmosphere at the relevant frequencies. We parameterize the atmosphere by allowing the ammonia abundance to vary between various pre-set pressures until we reach the ammonia condensation pressure, above which the profile follows the saturated vapor pressure curve \citet[Model 2]{Moeckel2023}. Figure \ref{fig:PJ19-NH3-ZonalMap} shows the ammonia distribution based on the averaged PJ19 observations assuming the atmosphere follows a dry adiabat \citep{Moeckel2023} anchored to the Galileo 1 bar temperature of 166.1 K \citep{Seiff1996}. While in recent years there is growing evidence from theory \citep{Guillot1995, Leconte2017, Guillot2020}, from simulations \citep{Li2018,Ge2024}, and from observations \citep{Li2024,Moeckel2024} that there are temperature variations below the clouds, we still have little information about their distribution. Although the exact temperature structure will affect the reference atmosphere \citep{Li2024,Moeckel2024}, it will not affect the main conclusions of this paper, which is based on relative measurements in the brightness temperature. We used an optimizer that minimizes the difference between the observables (nadir brightness temperature $T_{B}$, limb-darkening coefficient $p$) and their modeled counterparts ($T_m, p_m$), while simultaneously regularizing the amount of vertical fluctuations using $\lambda = 1E4$ \citep{Moeckel2023}:

\begin{equation}\label{eq:C1}
    C_{fit} = \sum_{1}^{6} \frac{(T_{m} - T_{B})^{2}}{\sigma_{T_{B}}^{2}} +  \sum_{2}^{5} \frac{(p_{m} - p)^{2}}{\sigma_{p}^{2}} + \lambda  \mid \sum_{1}^{8}\left({ \frac{\delta NH_3}{\delta ln(P)}}\right)_i\mid
\end{equation}

Although working in a Markov Chain Monte Carlo framework \citep{Li2017,Li2024,Moeckel2024} is preferable to determine parameters and their uncertainties, this was not possible due to the large computational costs incurred. Instead, we retrieve the zonally averaged ammonia abundance with 0.5$^\circ$ latitudinal resolution over the areal extent covered by Juno. Since optimizers have a higher likelihood to find local minima and produce large outliers, we apply a running median filter that is 2$^\circ$ wide to smooth out the data. Overall, the structure for PJ19 mirrors the topology we have seen before \citep{Moeckel2023}. We find an atmosphere that appears to be depleted in ammonia down to 30-40 bar in the southern hemisphere. The equator and northern hemisphere are dominated by the EZ and NEB signal, which have the lowest and highest brightness temperatures across the planet. We find a strong equatorial enhancement of the ammonia abundance especially at the northern edge of the EZ, with a highly depleted NEB that shows a charactistic depletion down to 40 bars, much deeper when compared to the equivalent latitudes in the southern hemisphere.

Next, we aim at retrieving residuals in the model atmosphere that mirror the residuals seen across the Juno maps. The main disadvantage of this method over absolute retrievals is the fact that we lose information about the absolute abundance and our results are relative to a reference atmosphere. However, the advantage of a relative retrieval is that we can remove the impact of the absolute calibration of 1.5\% and instead use the $\sim$0.1\% measurement uncertainty for each observation  \citep{Janssen2017}. The spatial variations within the maps are not affected by the absolute calibration error, which affects all data equally.  By using this method, we drastically increase the reliability of our retrievals. 

\begin{figure}
\centering
\includegraphics[trim={0cm 0 0.1cm 0},clip,width=\textwidth]{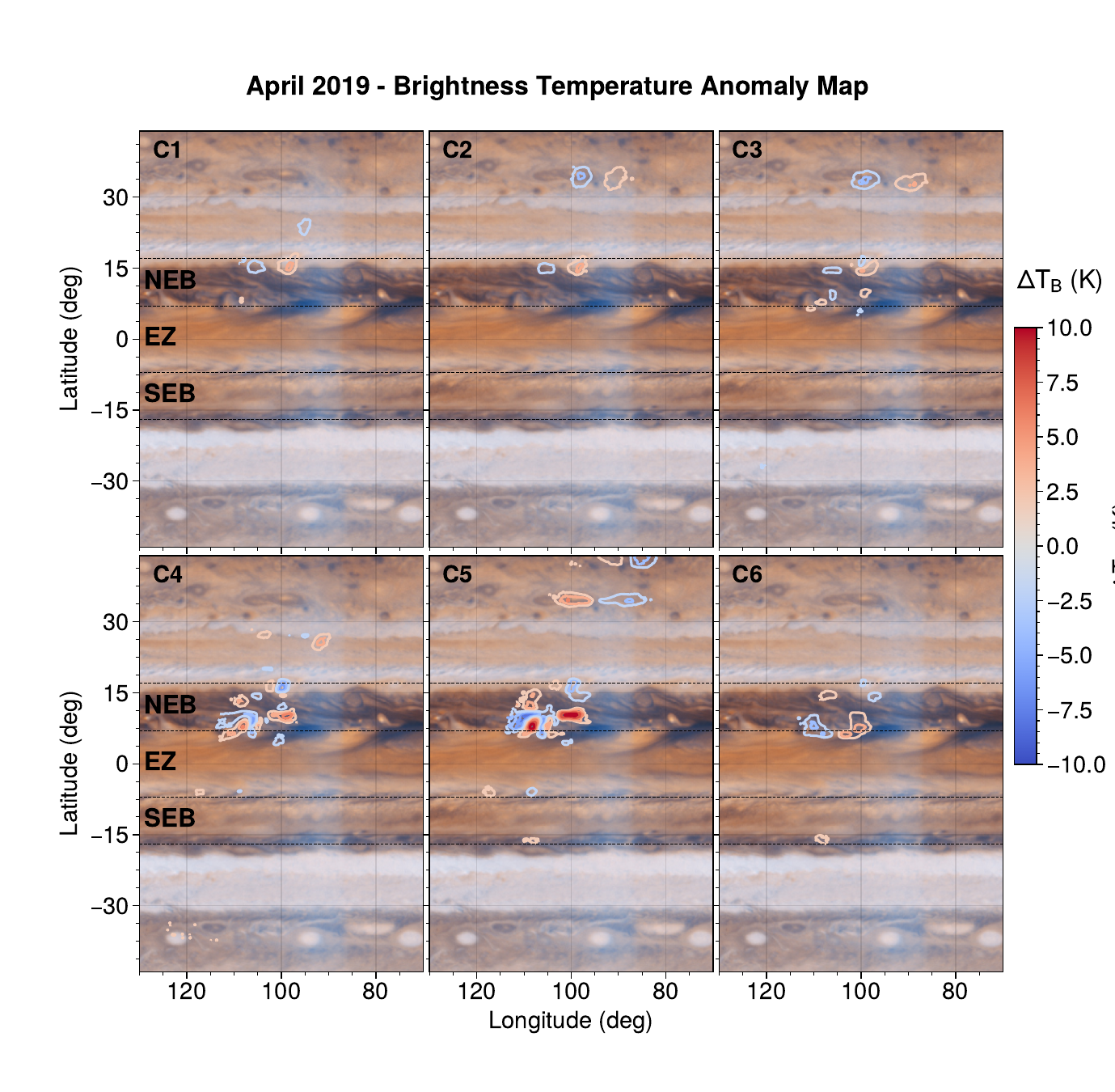}
\caption{Longitudinal brightness temperature anomalies after subtracting the zonal atmospheric brightness temperature at each frequency. The contours show temperature anomalies that are larger than 2K for both cold and warm anomalies. }
\label{fig:PJ19-AMap-Panel_v12_cleaned}
\end{figure}

To produce the anomaly map ($\Delta T(\theta,\phi)$), we first subtract the latitudinal (i.e., zonally averaged) brightness temperature profile ($\overline{T_B}(\theta)$) from the brightness temperature maps ($T(\theta,\phi)$ in Figure \ref{fig:juno-maps}) to produce the residual map (Figure \ref{fig:PJ19-AMap-Panel_v12_cleaned}): 

\begin{equation}
    \Delta T_B(\theta,\phi) = T_B(\theta,\phi) - \overline{T_B}(\theta)
\end{equation}

Additionally, we apply the synchrotron filter for Channel 1, 2, and 3 as explained in Appendix A to remove the mild limb-brightening in the residuals.

Using this framework, we retrieve an atmospheric model that shows the same residual structure as the observations. That is, we are no longer retrieving absolute ammonia abundances, but rather the ammonia anomaly with respect to our reference atmosphere, which would produce the same magnitude anomaly as we see in the observations.  We note that the variations in the atmosphere could be caused by either temperature or ammonia, however the regions over which the atmosphere is affected is the same regardless of which quantity is modeled \citep{Moeckel2024}. With this caveat addressed, we employ the following cost-function for the optimizer: 

\begin{equation}
    \begin{split} 
        C_{fit} = & \left(\sum_{1}^{6}  \frac{\overbrace{(T_{PJ19} - \bar{T})^{2}}^{Model} -  \overbrace{(T_{PJ19} - \bar{T})^{2}}^{Observations}}{\sigma_T^{2}} \right) \\ &+ \lambda  \mid \sum_{1}^{8}\left({ \frac{\delta NH_3}{\delta ln(P)}}\right)_i\mid
    \end{split} 
    \label{eq:CF_rel}
\end{equation}

We sampled the anomaly map at 0.5$^\circ$ resolution between -20$^\circ$ and 20$^\circ$ latitude. For each pixel, we fit the cost function in Equation \ref{eq:CF_rel}. In total, we used the retrieval on $\sim$ 3500 individual locations on the planet to build up a three-dimensional structure of the tropical subcloud atmosphere on Jupiter.

Higher up in the atmosphere, in Channels 4-6, we see significant variations in the brightness temperature, much greater than the expected noise floor of the instruments, resulting in strong constraints on the NH$_3$ abundance retrieved values. Deeper in the atmosphere, the residuals decrease in magnitude and the measurement error increases, making the retrieval at these pressures less certain. We employ a median filter at constant pressure that operates on a 1.5$^{\circ}$x1.5$^{\circ}$ grid to reduce the impact of outliers on the atmosphere.

\section{Modeling Results}\label{sec:results} 

For our analysis of the subcloud structure of Jupiter, we make an important distinction between the climatological signal and the weather. Circulation cells tend to transport material across latitudes, and thus the latitudinal distribution, as shown in Figure \ref{fig:PJ19-NH3-ZonalMap}, is composed of both small-scale weather effects, and the large-scale background circulation. In contrast, the distribution as a function of longitude is more indicative of local weather events, so for the purpose of this work, we refer to the longitudinal variations as weather. 

VLA observations cover the pressure regions around and below the ammonia ice cloud (see Figure \ref{fig:PJ19WF}), with the K-band frequencies (18-26 GHz) probing the saturation vapor curve inside and above the ammonia ice cloud, and Ku- and X-band observations covering the subcloud structure. Comparing the distribution between the various VLA frequencies, we can already glean some important information about the weather patterns on Jupiter. The red colors in the radio-composite image (see Figure \ref{fig:radio-composite}) show variations around a few bars. The overall lack of red and yellow hues indicates that the majority of the atmospheric variability is happening around and below the ammonia clouds, but does not extend to much greater depths. Most of the cloud dynamics are surprisingly shallow. There are several features, however, that clearly show up in the red color scheme and can also be distinguished in the X-band maps themselves. The NEB and SEB both show up as distinct features that wrap around the planet, and embedded within these belts we see variations. The NEB is disrupted by the repeated pattern of the 5 micron hotspots, and cold ammonia plumes that shape much of the interface between the EZ and NEB \citep{dePater2016}. The SEB is dominated by the Great Red Spot (GRS); especially its wake and bright ring around the GRS extend to several bars at least based on the VLA data \citep{dePater2019b}. The most consistent signal is a narrow band of warm brightness temperature at $\sim$15$^\circ$S, with variations that correlate well with the location of vortices on the southern edge of the SEB. The last category of features that appears to have large vertical extent in the atmosphere of Jupiter are the barges (the remnants of past storms \citep{Hueso2022}), such as the one just west of Oval BA in the southern hemisphere. Note that Oval BA itself is visible in the K and Ku bands, but not at X band. Similarly, longitude extended features can be identified in the northern hemisphere in the radio images (e.g. 140$^\circ$-125$^\circ$W, 40$^\circ$N,  or 70$^\circ$-60$^\circ$W, 40$^\circ$N). In the HST maps, these features appear as regions with a lot of substructure and a higher degree of turbulence, similar to cyclonic features, as opposed to the more continuous cloud deck seen at other longitudes.

The visible images and the brightness distribution maps (Figures \ref{fig:xku-maps} and \ref{fig:k-maps}) are anticorrelated because of the different sources of emission. Radio-cold regions are enhanced in traces gases, where the enhanced gases condense into clouds, which appear bright in the visible due to sunlight scattered off the clouds. In contrast, regions that are darker in the visible indicate low cloud coverage, so that  light penetrates deeper before it is scattered. The low cloud coverage also implies less trace gases and thus radio-warm regions.


Next, we focus on the tropical region that was sampled by MWR. From the longitudinal brightness temperature anomaly maps in Figure \ref{fig:PJ19-AMap-Panel_v12_cleaned} we can already discern some interesting features. The southern hemisphere shows much less spatial variability compared to the northern hemisphere.
The SEB shows a surprising lack of structure, with only very few shallow features (i.e., only visible at high frequencies) at the edges of the SEB, such as the feature on the southern edge around 110$^\circ$W, 16$^\circ$S and at the northern edge around 115$^\circ$W, 7$^\circ$S. 

The equatorial structure is also remarkably homogeneous, showing no significant variations with longitude, with the only exception of a localized cold feature at the northern edge of the EZ around 100$^\circ$W, 6$^\circ$N. This localized cold feature can be traced from Channel 3 to Channel 5.

The most significant variability is seen in the NEB with the strongest anomalies in Channel 5, i.e., just below the ammonia ice clouds. The majority of the signal is contained in the upper 4 channels but decreases deeper in the planet. 
North of the NEB, the spatial variability is again reduced with the exception of two distinct characteristics that are present in the northern hemisphere. There is a vortex at 15$^\circ$N at the boundary of NEB and NTrZ, and a barge-like feature at 34$^\circ$N \citep{Bolton2021}. The vortex can be identified across all 6 channels, indicating that it has a substantial depth to it, while the barge-like features only appears in the upper 5 channels.

\begin{figure}
\centering
\includegraphics[width=\textwidth]{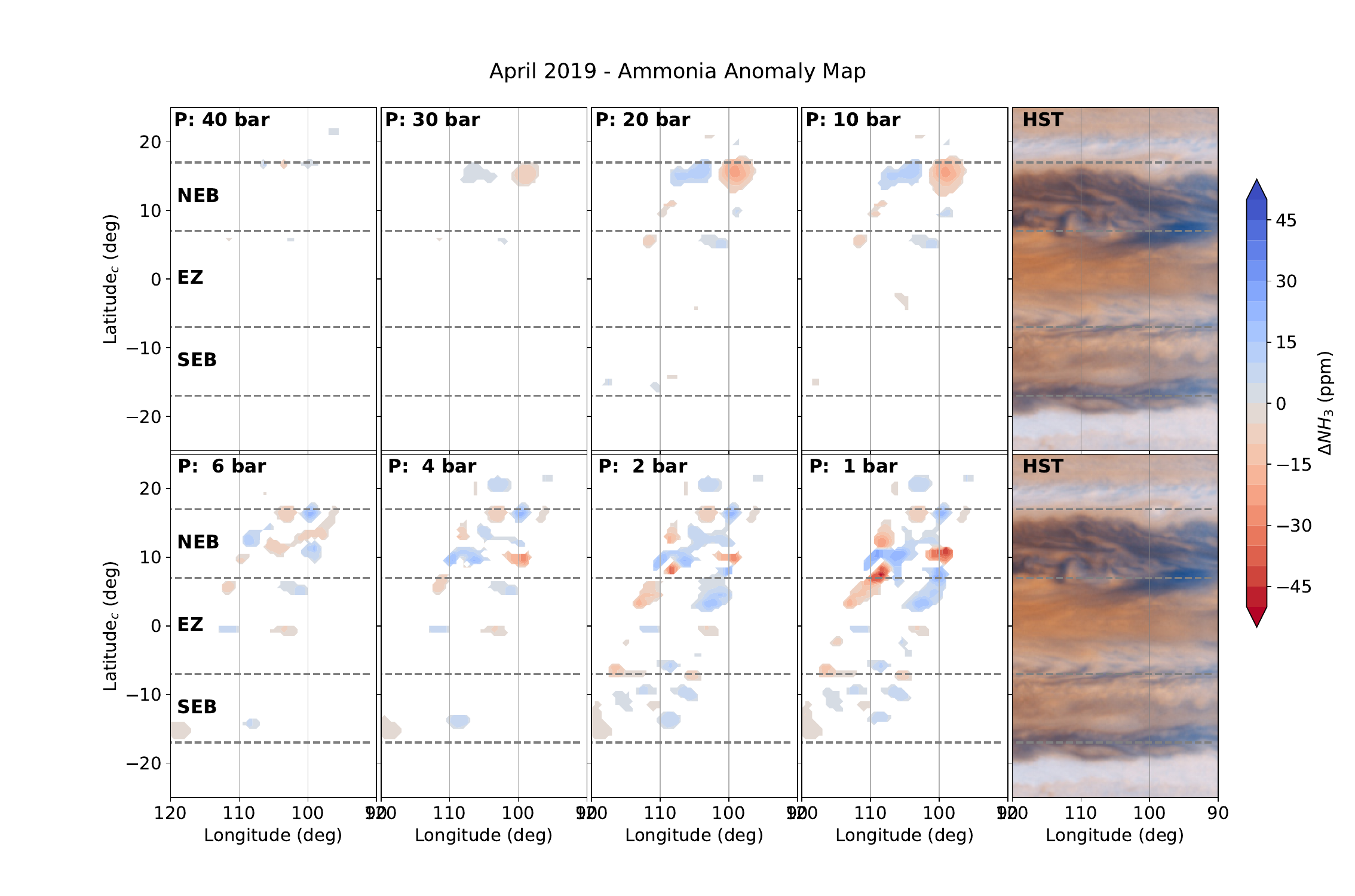}
\caption{Ammonia abundance anomalies that fit the brightness temperature anomalies in Figure \ref{fig:PJ19-AMap-Panel_v12_cleaned}. The right-most panel shows the (near)-simultaneous HST observations obtained within 2 hours of the flyby, and the panels to the left show slices in pressure, with the pressure indicated in the upper left corner. The ammonia variability peaks just below the clouds, and decreases with depth. Most features extend down to about 6 bars, where the water is expected to condense out, with the anticylone in the northern hemisphere and the ammonia plume at the interface between the NEB and EZ the two notable exceptions. The dynamics in the SEB and EZ appear to be mostly located at altitudes above 2 bar, while the NEB shows significant variations down to 6 bar. }
\label{fig:PJ19-DNH3-Map}
\end{figure}

Using Equation \ref{eq:CF_rel}, we can fit ammonia anomalies to the brightness temperature anomalies. Figure \ref{fig:PJ19-DNH3-Map} visualizes the retrieved ammonia anomaly at several pressures. The southern hemisphere and most of the equatorial zone, shows remarkably little variation, with most of the variability restricted to the upper 2 bars. The large residuals in the NEB can be traced down to the 6 bar levels.

Hence, overall, most of the variations in ammonia abundance, up to 50ppm, is near the 1 bar level, just under the ammonia ice clouds; the signal decays in magnitude with depth. Around the water condensation level ($\sim$6 bar), the variability has decreased considerably. There are a few exceptions to this trend: 

{\it i) Precipitation}
We find evidence for precipitation and evaporation in the form of dipolar structures in the ammonia anomalies. We find a depletion at low pressures high up in the atmosphere and an enhancement of ammonia below. For example  the depletion around (100$^\circ$W, 10$^\circ$N), turns into an enhancement around 6 bar, indicative of material being rained out and later evaporated to create the anomalies. 

{\it ii) Plume} 

An extended column of ammonia enhancement is seen at the interface of the NEB and EZ, that appears just west of the dark region in the HST images. These dark regions in HST maps are usually associated with hot spots \citep{Allison1990,Friedson2005}. The VLA X-band maps that provide context for these observations show an extended cold region east of the flyby, just overlapping with the edge of the MWR domain. Both these observations taken together are a strong indicator that it is the location of an ammonia plume as suggested before by \citet{Wong2020}. The MWR observations are able to trace this ammonia plume to about 30 bar. 

{\it iii) Vortex} 
Another interesting structure in the north tropical region is the vortex at 17$^\circ$N, at the interface of the NEB and North Tropical Zone (NTrZ). We find that higher up in the atmosphere the ammonia is enhanced causing the cold brightness temperature anomaly. Around the pressure where the water is condensing ($\sim$6 bar), the anomalies change sign and can be traced down to 40 bars. This finding is in line with the finding by \citet{Bolton2021} for the depth of the anticylone; however, we can rule out variations that go much deeper than 40 bar.  Additionally the ammonia depletion below the vortex may be shifted towards the south. As shown in Figure \ref{fig:PJ19WF}, most of the signal at $P>10$ bar originates from Channel 1 and 2, which have about twice the beamsize compared to the upper four channels. Therefore, the increased size of the anomaly can at least partially be explained by resolution effects. 
\\

\begin{figure}
\centering
\includegraphics[trim={23cm 0 8cm 0},clip,width=\textwidth]{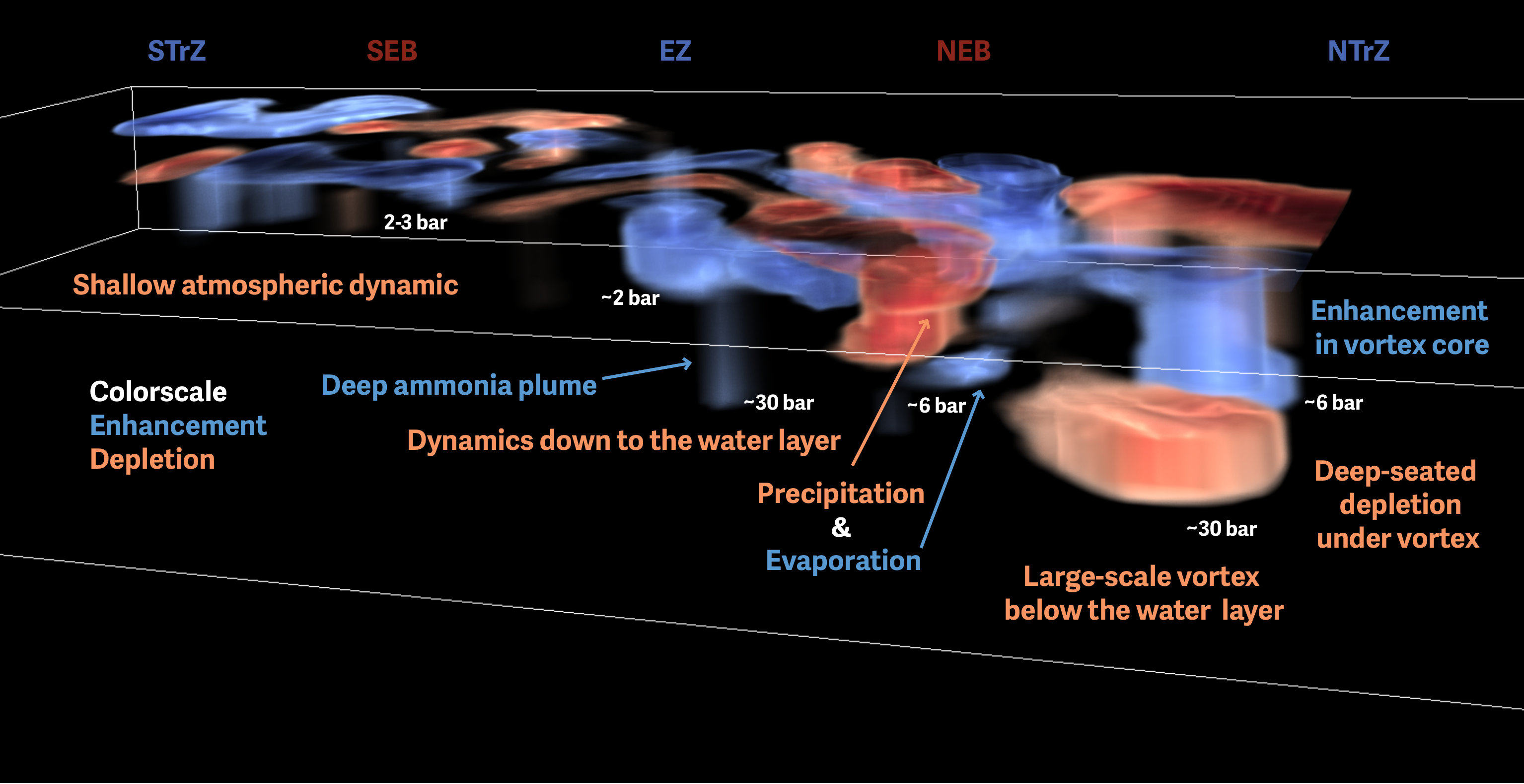}
\caption{3D rendering of the longitudinal ammonia anomaly (Figure \ref{fig:PJ19-DNH3-Map}) in the Jovian tropics, where red shades are ammonia depletion causing warm brightness temperature anomalies, while blue shades correspond to ammonia enhancements. The relative shallow atmospheric variations in the SEB are shown in the upper left corner, the EZ in the center, and the NEB and the vortex at its northern boundary in the bottom right corner. Processes that shape the atmosphere are highlighted, and approximate pressure levels are indicated. }
\label{fig:PJ19-render}
\end{figure}

To further illustrate the dynamics in the tropical region we render the anomaly with the "Visualization and Analysis Platform for Ocean, Atmosphere, and Solar Researchers" (VAPOR) \citep{Li2019V}. Figure \ref{fig:PJ19-render} shows a visualization of the anomaly, where blue regions indicate ammonia enhancement, and the orange / red shades show ammonia depletion. On the left-hand side, the SEB and EZ show very little spatial variability, except for the ammonia plume. As we approach the NEB the ammonia variations increase drastically. In this region there is evidence for precipitation in the form of depletion higher up in the atmosphere as ammonia is being removed, and the subsequent evaporation and enhancement around the water condensation level. At the northern edge of the rendering, the dipolar vortex structure is apparent, characterized by an enhanced ammonia core above a deep-seated ammonia depletion. 

\section{Discussion}\label{sec:discussion} 
The ammonia anomaly distribution discussed above gives us a first glimpse of the effect of weather on Jupiter. There are several processes that can modify the ammonia distribution, such as the background circulation \citep{Showman2005,Duer2021}, to local dynamical effects \citep{Ingersoll2004,dePater2016,Ingersoll2017,Ge2024}, microphysics \citep{Sugiyama2011,Sugiyama2014,Guillot2020} and temperature effects \citep{Palotai2014}. Regardless of the mechanisms modulating the ammonia distribution, we find some remarkable weather signatures in the atmosphere of Jupiter. Both the SEB and EZ appear well-mixed with shallow weather signatures that only extend down to about ~2 bar, indicating that most of the ammonia variability is contained within the region where ammonia should condense, precipitate and evaporate \citep{Ingersoll2017}. The difference in dynamical activity between the NEB and SEB has been observed in the brightness temperature distribution of VLA and MWR obervations before \citep{Hardesty2025}, but we extend this analysis by showing that the activity in the EZ and SEB is really limited to the upper few bars of pressures. The EZ is surprisingly homogeneous and very similar to the SEB in terms of activity. When looking at the general activity in the EZ in the X-band images, we see that much of the dynamical activity is driven by strong ammonia plumes, which show up predominantly at the lower frequencies (C-band, X-band) but not at the higher frequencies \citep{dePater2016}. 

\begin{figure}
\centering
\includegraphics[width=\textwidth]{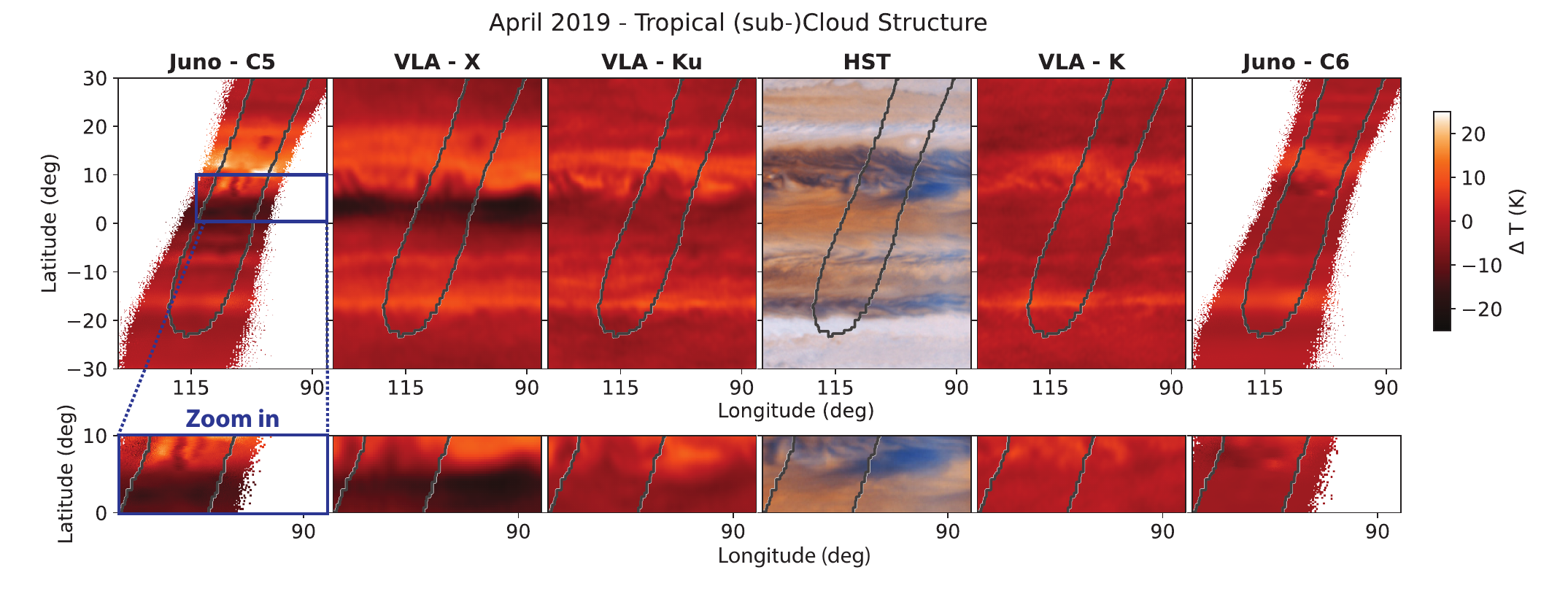}
\caption{Zoom in on the cloud and sub-cloud structure captured during the flyby. Comparing Channel 5 with the VLA observations that probe the same depths we can see that Juno passed just east of an extended cold region that appears to be part of the plume-hotspot system at the northern edge of the EZ. Furthermore we can see that Juno flew over a particularly variable region in the NEB. The VLA observations also show that variability indeed peaks around 1 bar, with observations probing a bit higher in the atmosphere showing much less contrast (Ku-, and K-band).}
\label{fig:PJ19-TropicalZoomMap}
\end{figure}

An intriguing feature of the MWR inversion is a deep-rooted ammonia enhancement of a few parts per million at the east end of the domain (101$^\circ$W, 6$^\circ$N). While the interpretation of the brightness temperature favored the shallow nature of the feature, we can trace the feature down to at about 30 bars, and it is the only feature that does not appear to change signs around the water condensation level. The context maps provided by the VLA in Figure \ref{fig:PJ19-TropicalZoomMap} show that the C5 domain appears to grace the tail end of an extended cold region, which we associate with ammonia plumes \citep{dePater2016}. We can clearly distinguish the plumes in the X-band observations; however, higher up in the atmosphere their signal fades away in the Ku- and K-band maps, which suggests that these are deep-seated features that have significant vertical structure and horizontal extent. Looking at the hotspot-plume cloud top manifestation (HST data) in Figure \ref{fig:PJ19-TropicalZoomMap}, we see a dark region at the cloud top (96$^\circ$W, 6$^\circ$N), and light-brown region east of that (90$^\circ$W, 6$^\circ$N). Indeed, \citet{Wong2020} suggested the presence of an ammonia plume in the observations based on Gemini-Near Infrared Observations and the HST observations. The VLA observations show that the plume extends much further west than the cloud top observations reveal, and show that the plume extends into the MWR domain. 

This shows the crucial role that ammonia plumes play in enriching the northern part of the EZ in the atmosphere by dredging up material from well below the water cloud, as originally suggested by \citet{dePater2016} based upon radiative transfer simulations of VLA 4--10 GHz data. In contrast, \citet{Fletcher2020b} suggested a shallow nature of the plumes based upon MWR 0.6--22 GHz brightness temperatures.  We use ammonia variations derived from MWR data combined with radiative transfer calculations to show that the plumes have a deep root and could potentially explain why the northern part of the EZ is enhanced compared to the southern edge. 

Further north in the domain, we enter the NEB, the region that shows the strongest brightness temperature anomalies, both for the zonal (see Figure \ref{fig:juno-observations}) and for the meridional brightness temperature distribution (see Figure \ref{fig:PJ19-AMap-Panel_v12_cleaned}). Even in this highly dynamical region, we find that the major dynamical activity in the NEB happens within the cloud-forming region of the planet. Contrasts of up to 20K were observed in Channel 5 in the brightness temperature anomalies, and yet we can explain most of the signal by strong ammonia depletions and enhancements around 1-2 bar. Using the VLA observations (see Figures \ref{fig:xku-maps} and \ref{fig:PJ19-TropicalZoomMap}), we see that the structure  probed by Juno is among the most extremes for this latitude. This allows us to generalize these findings that even in the most turbulent regions of the NEB, the majority of the variability is restricted to the weather layer (upper 6 bar), as also suggested by \citet{dePater2016}. The VLA observations further show that the variability indeed peaks around 1 bar (C5/X-band), with the observation probing slightly higher in the atmosphere (Ku-band) already showing a much reduced brightness temperature contrast.

The only small-scale structure that we see in the NEB that extends below the water layer appears to switch signs below the 4 bar level, similar to the signature seen during the SEB outbreak \citep{Moeckel2024}. Ammonia condenses above 0.7 bar level and precipitates in the form of ammonia rain. The depth to which a droplet can fall depends on its size and the atmospheric conditions \citep{Loftus2021}, but around 4 bar the ammonia is expected to boil \citep{Li2019}. This is the pressure at which we see the ammonia depletion at100$^\circ$W and 10$^\circ$N to turn into an enhancement consistent with the evaporation of ammonia droplets. However, the enhancements in ammonia can be traced to great pressures. This requires larger precipitates to penetrate through the water condensation layer \citep{Sugiyama2014, Guillot2020} and subsequent cold downdrafts to reach deeper levels \citep{Markham2023}.

With very few exceptions, we find that the majority of the weather on Jupiter is very shallow and that much of the brightness temperature variability in the upper channels from MWR and VLA observations can be explained by both local and global shallow atmospheric circulation models \citep{Showman2005,dePater2016,Ingersoll2017,Fletcher2020,Ge2024}. 

The depth of the vortex has already been established by \citet{Bolton2021}. Our results are mostly consistent with the finding that the vortex appears to have a denser upper core, above lighter/warmer region that appears to be shifted slightly southward compared to the upper core. 

The weather in the atmosphere appears too shallow to explain the depletion of ammonia down to 30-40 bar \citep{Li2017,Moeckel2023}. This requires processes that cause the global depletion of ammonia to either be much smaller than the resolution that we capture here \citep{Showman2005}, or require much more energy \citep{Li2023,Moeckel2024}.


\section{Conclusion}\label{sec:conclusions}
The distribution of ammonia anomalies offers valuable insights into Jupiter's atmospheric dynamics. By combining data from Juno's MWR with VLA and HST data, we conducted a groundbreaking study of Jupiter's tropical regions, generating for the first time a three-dimensional rendering of its weather system. We define weather as the longitudinal variations in the atmosphere after removing the latitudinal mean. Initially, we fitted a reference atmosphere based on various observables: nadir brightness temperature and limb-darkening effects. Subsequently, we extracted ammonia anomalies relative to this reference, enabling us to discern weather patterns beneath the cloud tops and investigate their vertical extent. 

Combining VLA and MWR observations, we conclude that Jupiter's weather is primarily confined to the uppermost layers of its atmosphere, suggesting a shallow weather system. Specifically, our analysis of MWR data revealed that while there are few variations within the North Equatorial Belt (NEB) that extend to below the water cloud, the majority of variability in the NEB occurs at altitudes above 6 bars, indicating that Jupiter's weather occurs predominantly between the water and ammonia cloud layers \citep{Showman2005,dePater2016,Fletcher2020,Ge2024}. The absence of significant structures at higher latitudes in both previous VLA maps \citep{dePater2019b} and in our VLA maps of this work are a strong indication that these results are valid beyond the tropics. Moreover, this underscores the critical role of water condensation in Jupiter's atmospheric dynamics, serving as a key energy source through latent heat release \citep{Ingersoll2000} and acting as a stabilizing agent to limit the dynamics to above the water layer  \citep{Guillot1995,Showman2005,Ge2024,Moeckel2024}.

Furthermore, we observed an ammonia enhancement at the NEB-EZ boundary, consistent with an ammonia plume reaching depths exceeding 30 bars — far deeper than previously interpreted \citep{Fletcher2020b}. This deep-rooted plume suggests continuous upwelling of material from the lower troposphere, explaining the ammonia enrichment in the equatorial zone compared to other regions \citep{Li2017,dePater2016,Moeckel2023}. Additionally, we identified an intriguing anticyclone at the NEB-NTrZ interface, characterized by a dense ammonia core above the water layer and ammonia depletion beneath it.

Our study underscores that conventional weather patterns observed in Jupiter's upper atmosphere do not fully account for the ammonia depletion down to 30 bars. Large-scale features such as plumes and vortices likely play pivotal roles in enriching and depleting Jupiter's atmosphere. Alternatively, the discrepancy may stem from storm-scale phenomena beyond the resolution of current observations \citep{Showman2005,Guillot2020}.

Future research, incorporating mid-infrared observations, holds promise for refining our understanding by resolving temperature ambiguities in the upper atmosphere. Additionally, integrating temperature structure retrievals, as proposed by \citet{Li2024}, could enhance the accuracy of the reference atmosphere, although the main findings regarding ammonia anomalies would remain unaffected.




\section*{Data Availability}
All data will be made available with the publication. The VLA data reduction, the MWR data reduction and deprojection code are all open-source and can be found on github (\url{https://github.com/cmoeckel91/pyPR}). The MWR data can be found on the Planetary Data System (\url{https://pds.nasa.gov/}). The HST observations can be found on the Mikulski Archive for Space Telescopes (\url{https://archive.stsci.edu/hlsp/wfcj}). 

\bibliography{main}{}

\begin{thebibliography}{}
\expandafter\ifx\csname natexlab\endcsname\relax\def\natexlab#1{#1}\fi
\providecommand{\url}[1]{\href{#1}{#1}}
\providecommand{\dodoi}[1]{doi:~\href{http://doi.org/#1}{\nolinkurl{#1}}}
\providecommand{\doeprint}[1]{\href{http://ascl.net/#1}{\nolinkurl{http://ascl.net/#1}}}
\providecommand{\doarXiv}[1]{\href{https://arxiv.org/abs/#1}{\nolinkurl{https://arxiv.org/abs/#1}}}

\bibitem[{{Allison}(1990)}]{Allison1990}
{Allison}, M. 1990, \icarus, 83, 282, \dodoi{10.1016/0019-1035(90)90069-L}

\bibitem[{{Bolton} {et~al.}(2021){Bolton}, {Levin}, {Guillot}, {Li}, {Kaspi}, {Orton}, {Wong}, {Oyafuso}, {Allison}, {Arballo}, {Atreya}, {Becker}, {Bloxham}, {Brown}, {Fletcher}, {Galanti}, {Gulkis}, {Janssen}, {Ingersoll}, {Lunine}, {Misra}, {Steffes}, {Stevenson}, {Waite}, {Yadav}, \& {Zhang}}]{Bolton2021}
{Bolton}, S.~J., {Levin}, S.~M., {Guillot}, T., {et~al.} 2021, Science, 374, 968, \dodoi{10.1126/science.abf1015}

\bibitem[{{de Pater} {et~al.}(2005){de Pater}, Deboer, Marley, Freedman, \& Young}]{dePater2005}
{de Pater}, I., Deboer, D., Marley, M., Freedman, R., \& Young, R. 2005, Icarus, 173, 425, \dodoi{10.1016/j.icarus.2004.06.019}

\bibitem[{{de Pater} {et~al.}(2023){de Pater}, {Molter}, \& {Moeckel}}]{depater2023}
{de Pater}, I., {Molter}, E.~M., \& {Moeckel}, C.~M. 2023, Remote Sensing, 15, 1313, \dodoi{10.3390/rs15051313}

\bibitem[{{de Pater} {et~al.}(2016){de Pater}, {Sault}, {Butler}, {DeBoer}, \& {Wong}}]{dePater2016}
{de Pater}, I., {Sault}, R.~J., {Butler}, B., {DeBoer}, D., \& {Wong}, M.~H. 2016, Science, 352, 1198, \dodoi{10.1126/science.aaf2210}

\bibitem[{{de Pater} {et~al.}(2019{\natexlab{a}}){de Pater}, {Sault}, {Wong}, {Fletcher}, {DeBoer}, \& {Butler}}]{dePater2019b}
{de Pater}, I., {Sault}, R.~J., {Wong}, M.~H., {et~al.} 2019{\natexlab{a}}, Icarus, 322, 168, \dodoi{10.1016/j.icarus.2018.11.024}

\bibitem[{{de Pater} {et~al.}(2014){de Pater}, {Fletcher}, {Luszcz-Cook}, {DeBoer}, {Butler}, {Hammel}, {Sitko}, {Orton}, \& {Marcus}}]{dePater2014}
{de Pater}, I., {Fletcher}, L.~N., {Luszcz-Cook}, S., {et~al.} 2014, \icarus, 237, 211, \dodoi{10.1016/j.icarus.2014.02.030}

\bibitem[{{de Pater} {et~al.}(2019{\natexlab{b}}){de Pater}, Sault, Moeckel, Moullet, Wong, Goullaud, Deboer, Butler, Bjoraker, {\'{A}}d{\'{a}}mkovics, Cosentino, Donnelly, Fletcher, Kasaba, Orton, Rogers, Sinclair, \& Villard}]{dePater2019}
{de Pater}, I., Sault, R.~J., Moeckel, C., {et~al.} 2019{\natexlab{b}}, The Astronomical Journal, 158, 139, \dodoi{10.3847/1538-3881/ab3643}

\bibitem[{{Duer} {et~al.}(2021){Duer}, {Gavriel}, {Galanti}, {Kaspi}, {Fletcher}, {Guillot}, {Bolton}, {Levin}, {Atreya}, {Grassi}, {Ingersoll}, {Li}, {Li}, {Lunine}, {Orton}, {Oyafuso}, \& {Waite}}]{Duer2021}
{Duer}, K., {Gavriel}, N., {Galanti}, E., {et~al.} 2021, \grl, 48, e95651, \dodoi{10.1029/2021GL095651}

\bibitem[{{Fletcher} {et~al.}(2020{\natexlab{a}}){Fletcher}, {Kaspi}, {Guillot}, \& {Showman}}]{Fletcher2020}
{Fletcher}, L.~N., {Kaspi}, Y., {Guillot}, T., \& {Showman}, A.~P. 2020{\natexlab{a}}, \ssr, 216, 30, \dodoi{10.1007/s11214-019-0631-9}

\bibitem[{{Fletcher} {et~al.}(2009){Fletcher}, {Orton}, {Yanamandra-Fisher}, {Fisher}, {Parrish}, \& {Irwin}}]{Fletcher2009}
{Fletcher}, L.~N., {Orton}, G.~S., {Yanamandra-Fisher}, P., {et~al.} 2009, Icarus, 200, 154, \dodoi{10.1016/j.icarus.2008.11.019}

\bibitem[{{Fletcher} {et~al.}(2020{\natexlab{b}}){Fletcher}, {Orton}, {Greathouse}, {Rogers}, {Zhang}, {Oyafuso}, {Eichst{\"a}dt}, {Melin}, {Li}, {Levin}, {Bolton}, {Janssen}, {Mettig}, {Grassi}, {Mura}, \& {Adriani}}]{Fletcher2020b}
{Fletcher}, L.~N., {Orton}, G.~S., {Greathouse}, T.~K., {et~al.} 2020{\natexlab{b}}, Journal of Geophysical Research (Planets), 125, e06399, \dodoi{10.1029/2020JE00639910.1002/essoar.10502118.1}

\bibitem[{{Friedson}(2005)}]{Friedson2005}
{Friedson}, A.~J. 2005, \icarus, 177, 1, \dodoi{10.1016/j.icarus.2005.03.004}

\bibitem[{{Ge} {et~al.}(2023){Ge}, {Li}, \& {Zhang}}]{Ge2024}
{Ge}, H., {Li}, C., \& {Zhang}, X. 2023, In review - Nature Astronomy, 7

\bibitem[{{Guillot}(1995)}]{Guillot1995}
{Guillot}, T. 1995, Science, 269, 1697, \dodoi{10.1126/science.7569896}

\bibitem[{{Guillot}(2021)}]{Guillot2021}
{Guillot}, T. 2021, in European Planetary Science Congress, EPSC2021--422, \dodoi{10.5194/epsc2021-422}

\bibitem[{Guillot {et~al.}(2020)Guillot, Stevenson, \& Becker}]{Guillot2020}
Guillot, T., Stevenson, D.~J., \& Becker, H.~N. 2020, Journal of Geophysical Research : Planets, 1, \dodoi{10.1029/2020JE006403}

\bibitem[{{Hardesty} {et~al.}(2025){Hardesty}, {Moeckel}, \& {de Pater}}]{Hardesty2025}
{Hardesty}, J., {Moeckel}, C., \& {de Pater}, I. 2025, \psj, 6, 50, \dodoi{10.3847/PSJ/ada428}

\bibitem[{{Hueso} {et~al.}(2022){Hueso}, {I{\~n}urrigarro}, {S{\'a}nchez-Lavega}, {Foster}, {Rogers}, {Orton}, {Hansen}, {Eichst{\"a}dt}, {Ordonez-Etxeberria}, {Rojas}, {Brueshaber}, {Sanz-Requena}, {P{\'e}rez-Hoyos}, {Wong}, {Momary}, {J{\'o}nsson}, {Antu{\~n}ano}, {Baines}, {Dahl}, {Mizumoto}, {Go}, \& {Anguiano-Arteaga}}]{Hueso2022}
{Hueso}, R., {I{\~n}urrigarro}, P., {S{\'a}nchez-Lavega}, A., {et~al.} 2022, \icarus, 380, 114994, \dodoi{10.1016/j.icarus.2022.114994}

\bibitem[{{Ingersoll} {et~al.}(2000){Ingersoll}, {Gierasch}, {Banfield}, {Vasavada}, \& {Galileo Imaging Team}}]{Ingersoll2000}
{Ingersoll}, A.~P., {Gierasch}, P.~J., {Banfield}, D., {Vasavada}, A.~R., \& {Galileo Imaging Team}. 2000, Nature, 403, 630, \dodoi{10.1038/35001021}

\bibitem[{{Ingersoll} {et~al.}(2004){Ingersoll}, {Dowling}, {Gierasch}, S., L., Agustin, P., A., \& R.}]{Ingersoll2004}
{Ingersoll}, A.~P., {Dowling}, T.~E., {Gierasch}, P.~J., {et~al.} 2004, in {Chapter 6: Dynamics of Jupiter's atmosphere in Jupiter The Planet, Satellites and Magnetosphere}

\bibitem[{{Ingersoll} {et~al.}(2017){Ingersoll}, {Adumitroaie}, {Allison}, {Atreya}, {Bellotti}, {Bolton}, {Brown}, {Gulkis}, {Janssen}, {Levin}, {Li}, {Li}, {Lunine}, {Orton}, {Oyafuso}, \& {Steffes}}]{Ingersoll2017}
{Ingersoll}, A.~P., {Adumitroaie}, V., {Allison}, M.~D., {et~al.} 2017, Geophysical Research Letters, 44, 7676, \dodoi{10.1002/2017GL074277}

\bibitem[{Janssen {et~al.}(2017)Janssen, Oswald, Brown, Gulkis, Levin, Bolton, Allison, Atreya, Gautier, Ingersoll, Lunine, Orton, Owen, Steffes, Adumitroaie, Bellotti, Jewell, Li, Li, Misra, Oyafuso, Santos-Costaz, Sarkissian, Williamson, Arballo, Kitiyakaral, Ulloa-Severino, Chen, Maiwald, Sahakian, Pingree, A., Mazer, Redick, Hodges, Hughes, Bedrosian, Dawson, Hatch, Russell, Chamberlain, Zawadskil, Khayatianl, Franklin, Conley, Kempenaar, Lool, Sunada, Vorperion, \& Wang}]{Janssen2017}
Janssen, M.~A., Oswald, J.~E., Brown, S.~T., {et~al.} 2017, Space Science Review, 213, 139, \dodoi{10.1007/s11214-017-0349-5}

\bibitem[{{Leconte} {et~al.}(2017){Leconte}, {Selsis}, {Hersant}, \& {Guillot}}]{Leconte2017}
{Leconte}, J., {Selsis}, F., {Hersant}, F., \& {Guillot}, T. 2017, Astronomy \& Astrophysics, 598, \dodoi{10.1051/0004-6361/201629140}

\bibitem[{{Li} \& {Chen}(2019)}]{Li2019}
{Li}, C., \& {Chen}, X. 2019, Astrophysical Journal Supplement Series, 240, 37, \dodoi{10.3847/1538-4365/aafdaa}

\bibitem[{{Li} {et~al.}(2023){Li}, {de Pater}, {Moeckel}, {Sault}, {Butler}, {deBoer}, \& {Zhang}}]{Li2023}
{Li}, C., {de Pater}, I., {Moeckel}, C., {et~al.} 2023, Science Advances, 9, \dodoi{10.1126/sciadv.adg9419}

\bibitem[{{Li} {et~al.}(2018){Li}, {Ingersoll}, \& {Oyafuso}}]{Li2018}
{Li}, C., {Ingersoll}, A.~P., \& {Oyafuso}, F. 2018, Journal of the Atmospheric Sciences, 75, 1063, \dodoi{10.1175/JAS-D-17-0257.1}

\bibitem[{Li {et~al.}(2017)Li, Ingersoll, Janssen, Levin, Bolton, Adumitroaie, Allison, Arballo, Bellotti, Brown, Ewald, Jewell, Misra, Orton, Oyafuso, Steffes, \& Williamson}]{Li2017}
Li, C., Ingersoll, A., Janssen, M., {et~al.} 2017, Geophysical Research Letters, 44, 5317, \dodoi{10.1002/2017GL073159}

\bibitem[{{Li} {et~al.}(2024){Li}, {Allison}, {Atreya}, {Brueshaber}, {Fletcher}, {Guillot}, {Li}, {Lunine}, {Miguel}, {Orton}, {Steffes}, {Waite}, {Wong}, {Levin}, \& {Bolton}}]{Li2024}
{Li}, C., {Allison}, M., {Atreya}, S., {et~al.} 2024, \icarus, 414, 116028, \dodoi{10.1016/j.icarus.2024.116028}

\bibitem[{{Li} {et~al.}(2019){Li}, {Jaroszynski}, {Pearse}, {Orf}, \& {Clyne}}]{Li2019V}
{Li}, S., {Jaroszynski}, S., {Pearse}, S., {Orf}, L., \& {Clyne}, J. 2019, Atmosphere, 10, 488, \dodoi{10.3390/atmos10090488}

\bibitem[{{Loftus} \& {Wordsworth}(2021)}]{Loftus2021}
{Loftus}, K., \& {Wordsworth}, R.~D. 2021, Journal of Geophysical Research (Planets), 126, \dodoi{10.1029/2020JE006653}

\bibitem[{{Markham} {et~al.}(2023){Markham}, {Guillot}, \& {Li}}]{Markham2023}
{Markham}, S., {Guillot}, T., \& {Li}, C. 2023, Astronomy \& Astrophysics, 674, \dodoi{10.1051/0004-6361/202245609}

\bibitem[{{Moeckel} {et~al.}(2024){Moeckel}, , {Ge}, \& {de Pater}}]{Moeckel2024}
{Moeckel}, C., , {Ge}, H., \& {de Pater}, I. 2024, In review - Science Advances

\bibitem[{{Moeckel} {et~al.}(2023){Moeckel}, {de Pater}, \& {DeBoer}}]{Moeckel2023}
{Moeckel}, C., {de Pater}, I., \& {DeBoer}, D. 2023, Planetary Science Journal, 4, 25, \dodoi{10.3847/PSJ/acaf6b}

\bibitem[{{Palotai} {et~al.}(2014){Palotai}, {Dowling}, \& {Fletcher}}]{Palotai2014}
{Palotai}, C., {Dowling}, T.~E., \& {Fletcher}, L.~N. 2014, \icarus, 232, 141, \dodoi{10.1016/j.icarus.2014.01.005}

\bibitem[{{Santos-Costa} {et~al.}(2017){Santos-Costa}, {Adumitroaie}, {Ingersoll}, {Gulkis}, {Janssen}, {Levin}, {Oyafuso}, {Brown}, {Williamson}, {Bolton}, \& {Connerney}}]{SantosCosta2017}
{Santos-Costa}, D., {Adumitroaie}, V., {Ingersoll}, A., {et~al.} 2017, Geophysical Research Letters, 44, 8676, \dodoi{10.1002/2017GL072836}

\bibitem[{{Sault} {et~al.}(2004){Sault}, {Engel}, \& {de Pater}}]{sault2004}
{Sault}, R.~J., {Engel}, C., \& {de Pater}, I. 2004, \icarus, 168, 336, \dodoi{10.1016/j.icarus.2003.11.014}

\bibitem[{{Sault} {et~al.}(1995){Sault}, {Teuben}, \& {Wright}}]{Sault1995}
{Sault}, R.~J., {Teuben}, P.~J., \& {Wright}, M.~C.~H. 1995, in Astronomical Society of the Pacific Conference Series, Vol.~77, Astronomical Data Analysis Software and Systems IV, ed. R.~A. {Shaw}, H.~E. {Payne}, \& J.~J.~E. {Hayes}, 433, \dodoi{10.48550/arXiv.astro-ph/0612759}

\bibitem[{Seiff {et~al.}(1996)Seiff, Kirk, Knight, Mihalov, Blanchard, Young, Schubert, Zahn, Lehmacher, Milos, \& Wang}]{Seiff1996}
Seiff, A., Kirk, D.~B., Knight, T. C.~D., {et~al.} 1996, Science

\bibitem[{{Showman} \& {de Pater}(2005)}]{Showman2005}
{Showman}, A.~P., \& {de Pater}, I. 2005, Icarus, 174, 192, \dodoi{10.1016/j.icarus.2004.10.004}

\bibitem[{{Sugiyama} {et~al.}(2014){Sugiyama}, {Nakajima}, {Odaka}, {Kuramoto}, \& {Hayashi}}]{Sugiyama2014}
{Sugiyama}, K., {Nakajima}, K., {Odaka}, M., {Kuramoto}, K., \& {Hayashi}, Y.~Y. 2014, Icarus, 229, 71, \dodoi{10.1016/j.icarus.2013.10.016}

\bibitem[{{Sugiyama} {et~al.}(2011){Sugiyama}, {Nakajima}, {Odaka}, {Ishiwatari}, {Kuramoto}, {Morikawa}, {Nishizawa}, {Takahashi}, \& {Hayashi}}]{Sugiyama2011}
{Sugiyama}, K., {Nakajima}, K., {Odaka}, M., {et~al.} 2011, Geophysical Research Letters, 38, \dodoi{10.1029/2011GL047878}

\bibitem[{{Wong} {et~al.}(2020){Wong}, {Simon}, {Tollefson}, {de Pater}, {Barnett}, {Hsu}, {Stephens}, {Orton}, {Fleming}, {Goullaud}, {Januszewski}, {Roman}, {Bjoraker}, {Atreya}, {Adriani}, \& {Fletcher}}]{Wong2020}
{Wong}, M.~H., {Simon}, A.~A., {Tollefson}, J.~W., {et~al.} 2020, \apjs, 247, 58, \dodoi{10.3847/1538-4365/ab775f}

\end{thebibliography}
\bibliographystyle{aasjournal}

\newpage
\section*{Appendix A}\label{sec:appA}
After analyzing the residuals, we observed a near-symmetric increase in brightness temperature towards the edges of the Juno domain in Channel 1. Despite our method being restricted to observations where 99\% of the emission originates from the planet, uncertainties in the beam shape may pick up extra-planetary emission from the synchrotron radiation belts. Given that non-thermal emission is several times stronger than the thermal emission from the planet for Channel 1 \citep{SantosCosta2017}, even minor uncertainties can manifest prominently.

During normal operations, Juno scans the planet from north to south, optimized to acquire multiple measurements for a given latitude at different emission angles. For PJ19, however, the spacecraft is scanning the planet from west to east, resulting in a distinct distribution where the emission angle varies with longitude. Larger emission angles correspond to the edges of the domain, while the smallest angles are found towards the map's center. Although synchrotron radiation is expected to vary with latitude, the effect should be relatively consistent within approximately 100$^\circ$. We, therefore, initially centered the observations around the minimum emission angle and then applied a median filter with a width of 20$^\circ$ to the data (twice the width of a typical zone or belt). Increasing the window size beyond this threshold would fail to capture variations in synchrotron strength with latitude, while reducing the window size risked removing genuine atmospheric structures.

\begin{figure}
\centering
\includegraphics[width=0.75\textwidth]{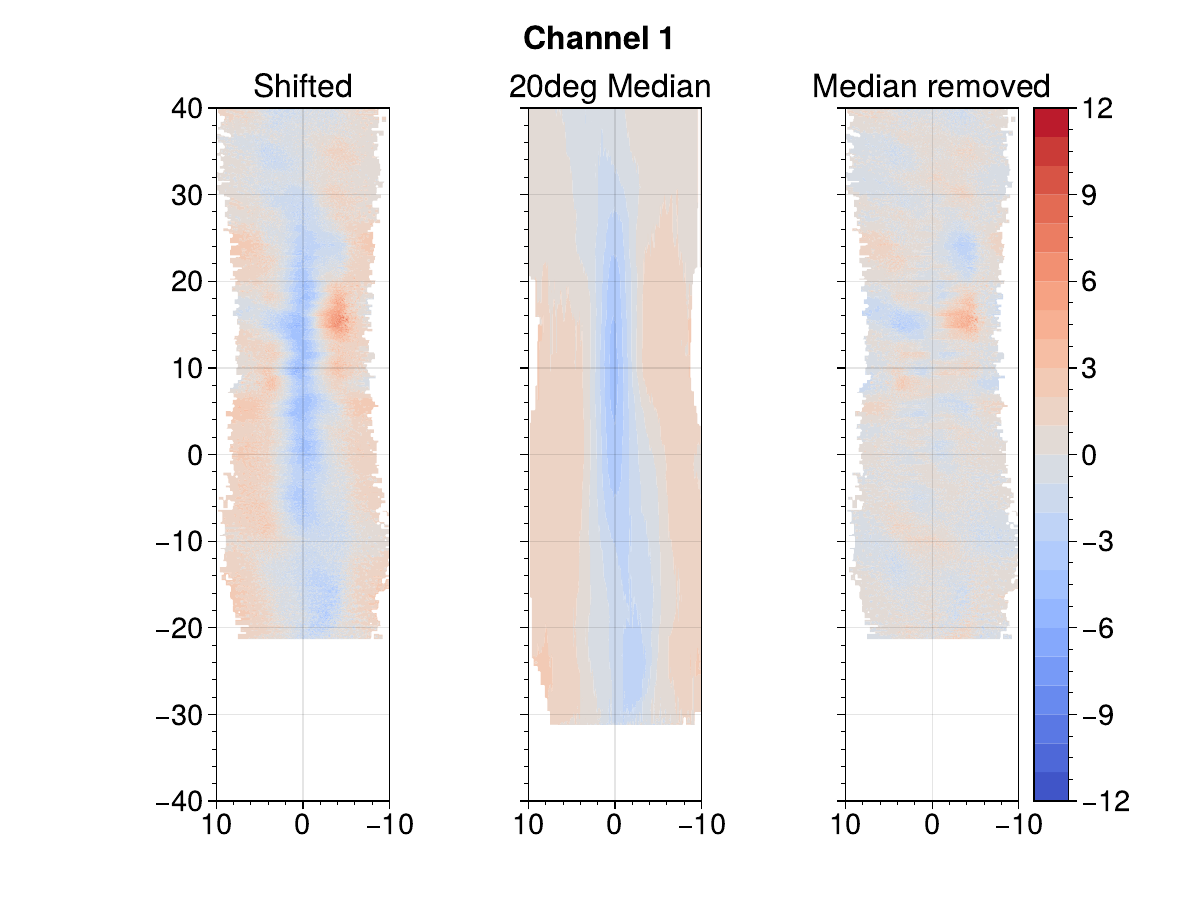}
\caption{The left panel displays the residuals after subtracting the mean of the PJ19 observations. We re-mapped the structure to be centered around the minimum emission angle to highlight its dependence on the emission angle and longitude. We calculated the running median over 20 degrees latitude to reveal the systematic brightening towards the limb, attributed to the increased synchrotron contribution. We chose a window size of 20 degrees to ensure it was twice as large as the typical extent of the zones and belts. In the right panel, the residual structure after removing the median is depicted. Although the majority of the atmospheric signal is attenuated, the structure at 15$^\circ$N persists, indicating a deep-seated origin for this feature.} 
\label{fig:AA-PJ19-TBMap_C1}
\end{figure}

The leftmost panel of Figure \ref{fig:AA-PJ19-TBMap_C1} displays the shifted data, revealing the residual temperature after subtracting the mean zonal brightness temperature. Notably, the increase in brightness temperature residual with distance from the map's center is evident in Channel 1. The center panel showcases the median-filtered residual, with the strongest residual observed around the equator, consistent with the distribution of synchrotron emission. Finally, the right panel displays the residual after removing the median-filtered data, where the deep-rooted vortex at 15$^\circ$N remains visible, indicating that genuine structure persists in our residual maps even after filtering. We applied a similar filter to Channels 2 and 3 but observed minimal effects on Channel 3, as expected due to the decreased strength of synchrotron emission with increasing frequency.

\begin{figure}
\centering
\includegraphics[width=0.75\textwidth]{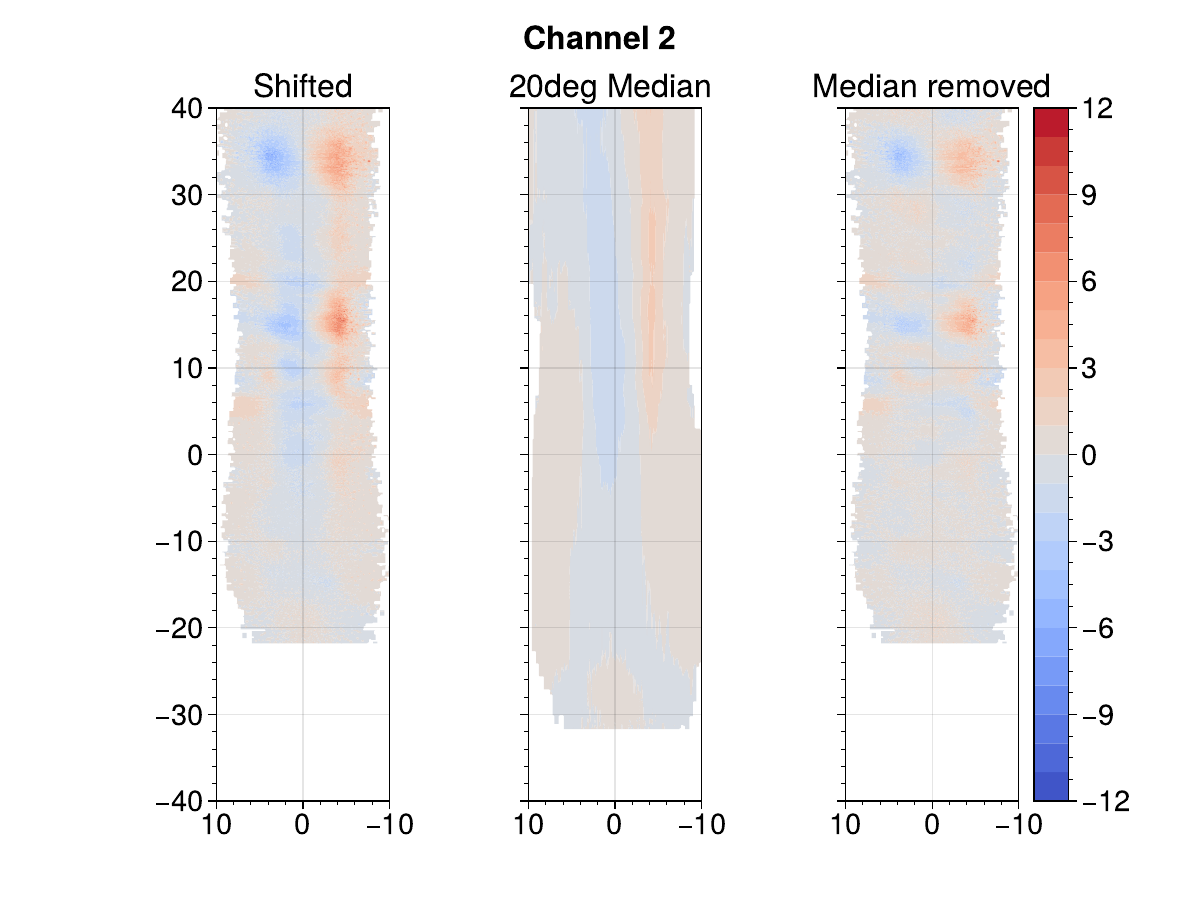}
\caption{Same as Figure \ref{fig:AA-PJ19-TBMap_C1} but for Channel 2 }
\label{fig:AA-PJ19-TBMap_C2}
\end{figure}

\begin{figure}
\centering
\includegraphics[width=0.75\textwidth]{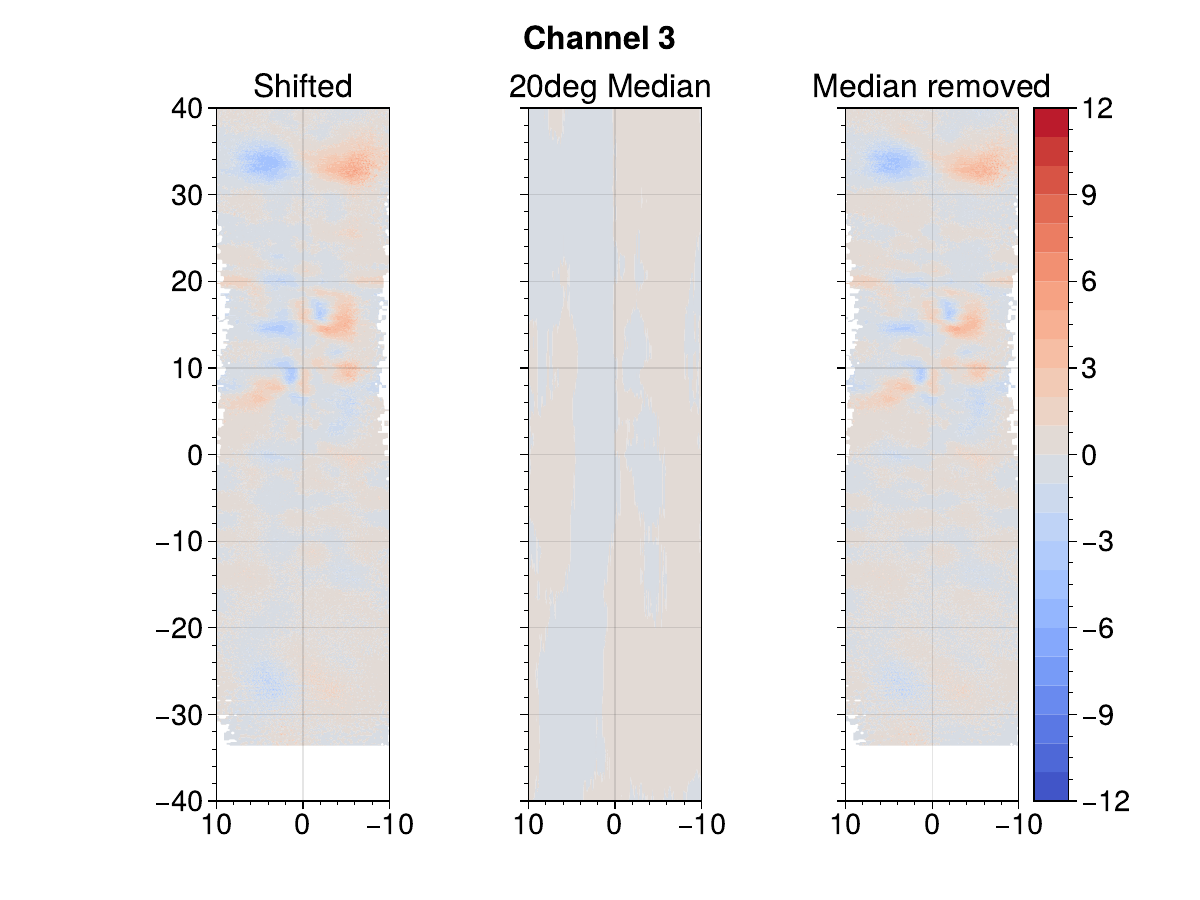}
\caption{Same as Figure \ref{fig:AA-PJ19-TBMap_C1} but for Channel 3}
\label{fig:AA-PJ19-TBMap_C3}
\end{figure}

\section*{Appendix B}\label{sec:appB}
Radio observations come with two inherent resolution limits. The largest separation between antennas indicates the highest achievable resolution for the radio maps, while the shortest baseline indicates the largest resolvable structure for radio observations. The VLA B-configuration strikes a good balance between these two resolution limits. However, after subtracting a limb-darkened disk, we identified large-scale structures in the VLA maps that seemed superimposed on the residual structure. This residual, on the order of a few Kelvins, manifested as a large sinusoidal structure particularly affecting the mid-latitudes. Since this large-scale structure must originate from observations at the shorter baselines, we restricted the mapping to observations above 15, 15, and 20 kilo$\lambda$ for the X-, Ku-, and K-band observations, respectively. Although we managed to reduce the magnitude of the residual, we were unable to fully eliminate it. The most likely cause is a slight misalignment between our model and the actual observations caused by the self-calibration processes. Instead, we applied a Gaussian filter (i.e., a high-pass filter) on the image plane, targeting the removal of the largest-scale structure. We selected a Gaussian kernel size that specifically targeted the large sinusoidal structure while preserving the smaller-scale structures of interest for our study. By subtracting the filtered image from the image plane, we reduced the amplitude of the residual, rendering our filtered maps an upper limit. The following images display the removed structure for the maps presented in Figures \ref{fig:xku-maps} and \ref{fig:k-maps}.

\begin{figure}
\centering
\includegraphics[width=0.9\textwidth]{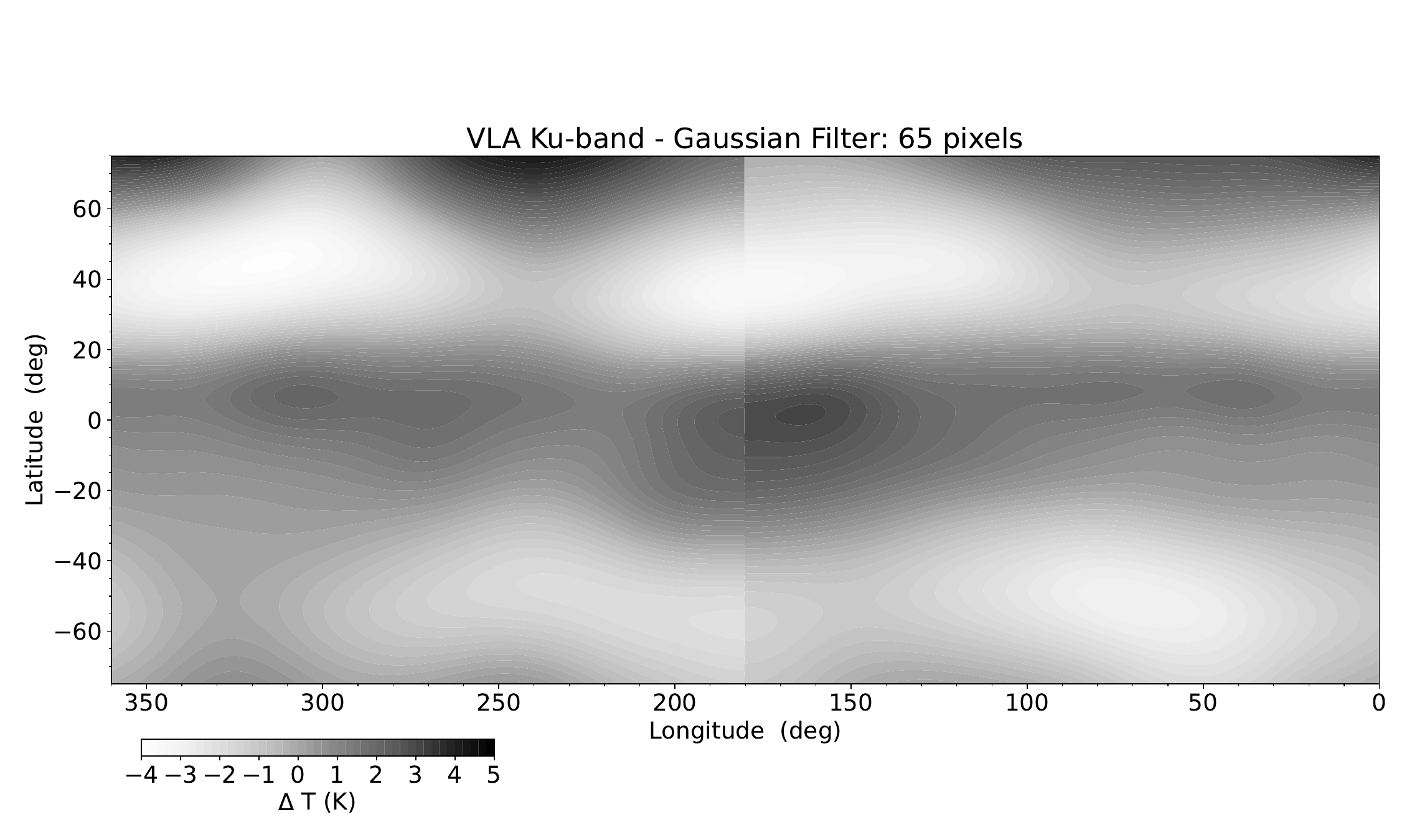}
\caption{Residual structure due to misalignment during the deprojection process that was removed for the Ku-band observations with a Gaussian filter of 65 pixels.} 
\label{fig:Ku_removed}
\end{figure}

\begin{figure}
\centering
\includegraphics[width=0.9\textwidth]{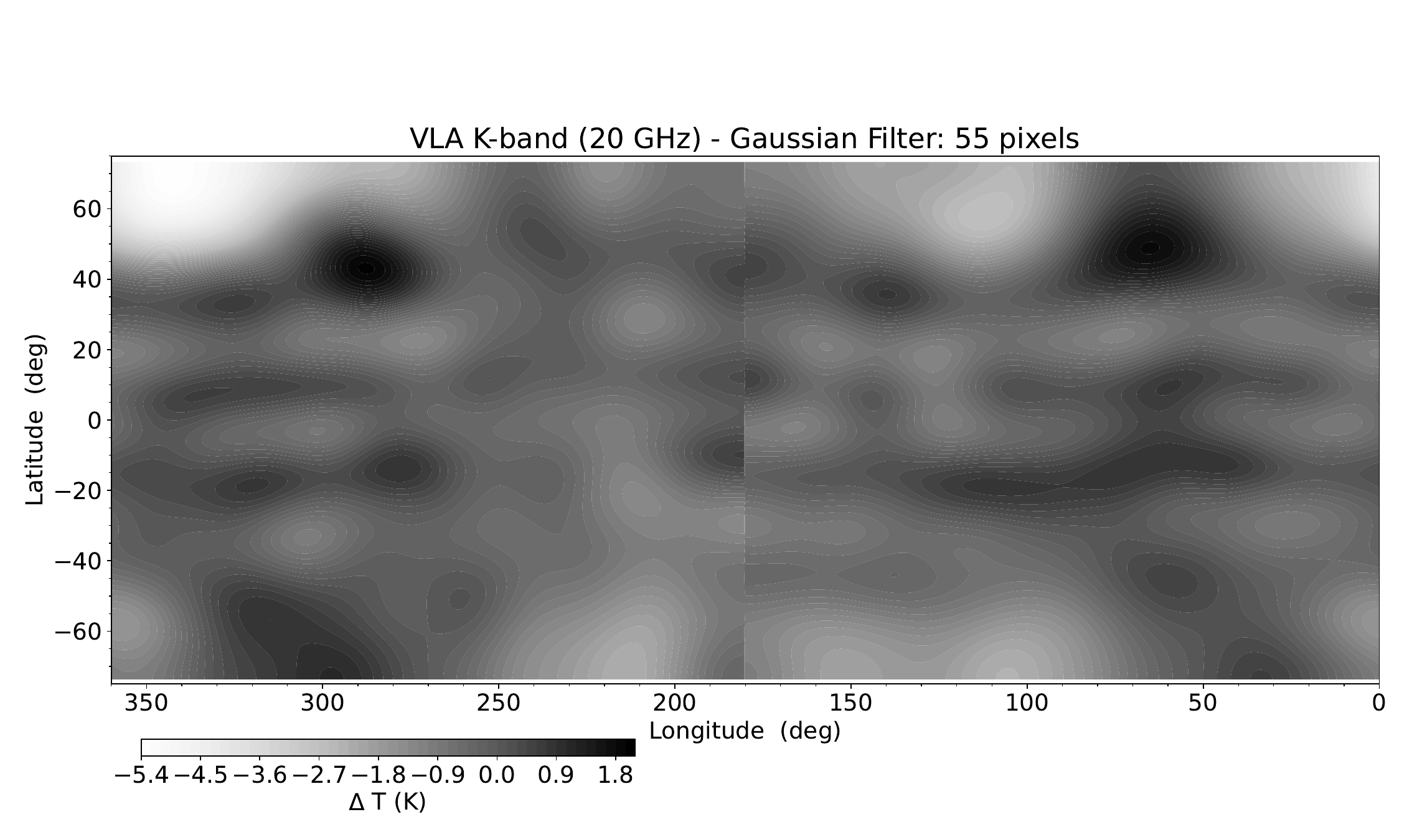}
\caption{Same as Figure \ref{fig:Ku_removed} but for K-band observations around 20 GHz.} 
\label{fig:K1_removed}
\end{figure}

\begin{figure}
\centering
\includegraphics[width=0.9\textwidth]{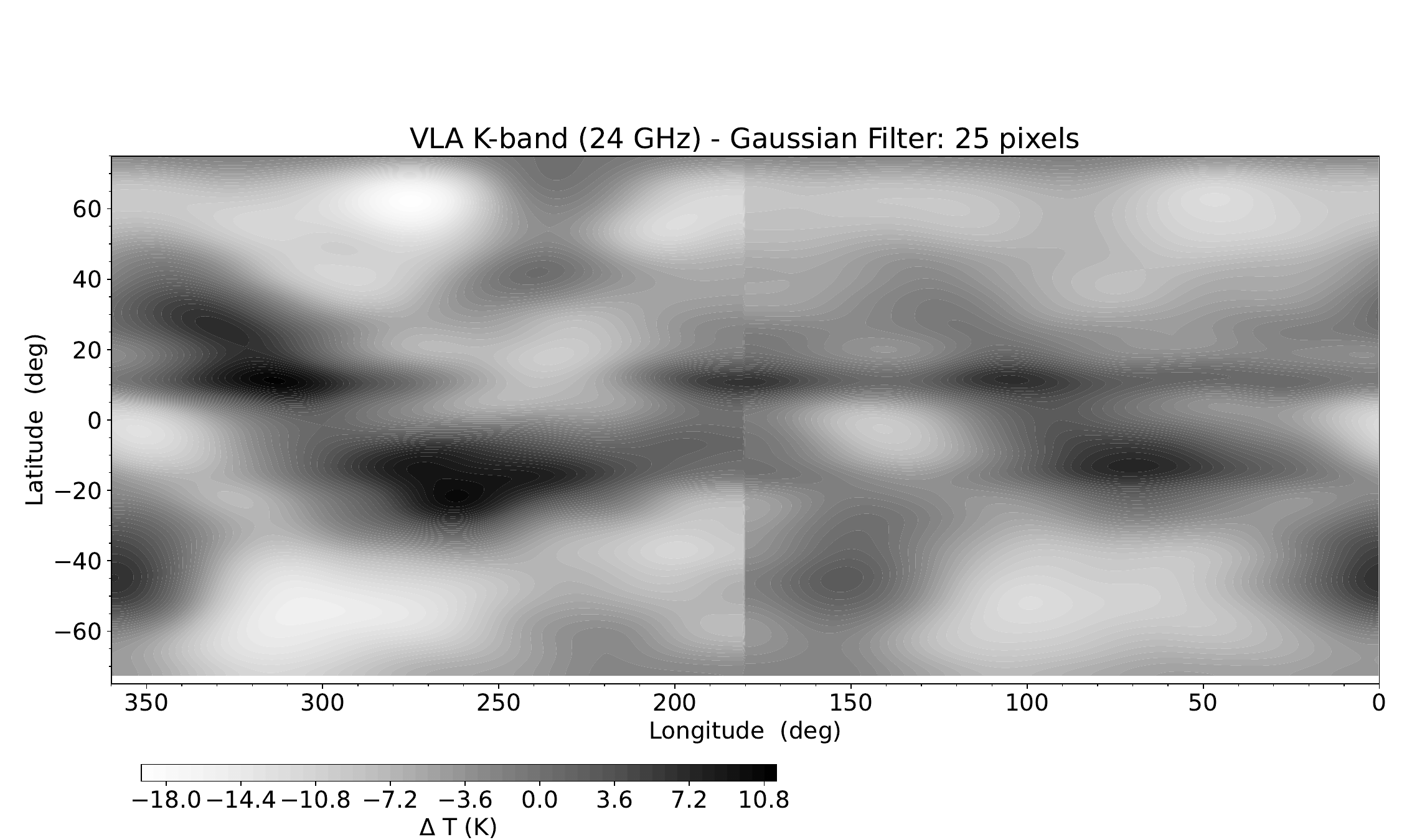}
\caption{Same as Figure \ref{fig:Ku_removed} but for K-band observations around 24 GHz.} 

\label{fig:K2_removed}
\end{figure}



\end{document}